\documentclass[10pt]{article}
\usepackage{mathrsfs}
\usepackage{amsmath,amsfonts,amssymb,mathtools}
\usepackage{cite}
\usepackage{dsfont}
\usepackage{graphicx}
\usepackage{hyperref}
\usepackage{verbatim}
\usepackage{slashed}
\usepackage[all]{xy}
\usepackage{xcolor}
\usepackage{subfig}
\usepackage{tikz}
\usetikzlibrary{shapes.geometric,arrows}
\usepackage{relsize}
\usepackage{caption}
\usepackage{float}
\usetikzlibrary{positioning}
\usepackage{geometry}
\usepackage[T1]{fontenc}
\usepackage{lmodern}
\usepackage{stackengine} %per \ocirc 
 
\usepackage[utf8]{inputenc}

\hypersetup{
  colorlinks,
  citecolor=violet,
  linkcolor=blue,
  urlcolor=blue}

\csname @addtoreset\endcsname{equation}{section}
\DeclareMathAlphabet\mathbfcal{OMS}{cmsy}{b}{n}

\newcommand{\beq}{\begin{equation}}
\newcommand{\eeq}{\end{equation}}
\newcommand{\bea}{\begin{eqnarray}}
\newcommand{\eea}{\end{eqnarray}}
\newcommand{\ba}{\begin{array}}
\newcommand{\ea}{\end{array}}
\newcommand{\bit}{\begin{itemize}}
\newcommand{\eit}{\end{itemize}}
\newcommand{\nn}{\nonumber}

\newcommand{\complesso}{{\ \hbox{{\rm I}\kern-.6em\hbox{\bf C}}}}
\newcommand{\reale}{{\hbox{{\rm I}\kern-.2em\hbox{\rm R}}}}
\newcommand{\Z}{\mathbb{Z}}  %  interi
\newcommand{\uno}{ \,  \raisebox{+0.14em}{{\hbox{{\rm \scriptsize ]}} \raisebox{-0.2em}{\kern-.8em\hbox{1}}}} \, }  %  operatore identit\`a

\newcommand{\p}{\partial}
\newcommand{\w}{\wedge}

\renewcommand{\a}{\alpha}
\renewcommand{\b}{\beta}
\newcommand{\g}{\gamma}

\newcommand{\G}{\Gamma}
\renewcommand{\d}{\delta}
\newcommand{\D}{\Delta}
\newcommand{\e}{\epsilon}
\newcommand{\Er}{{\mathbfcal{E}}}

\renewcommand{\k}{\kappa}
\renewcommand{\l}{\lambda}

\newcommand{\m}{\mu}

\newcommand{\n}{\nu}
\renewcommand{\r}{\rho}
\newcommand{\s}{\sigma}

\renewcommand{\S}{\Sigma}

\newcommand{\z}{\zeta}
\newcommand{\x}{\xi}

\newcommand{\om}{\omega}
\newcommand{\Om}{\Omega}
\newcommand{\y}{\psi}

\begin{document}

%\begin{comment}

\begin{titlepage}

\vspace{0.3cm}

\begin{flushright}
%$IFUM$--1105--$FT$ \\
$LIFT$--9-2.25
\end{flushright}

\vspace{0.3cm}

\begin{center}
\renewcommand{\thefootnote}{\fnsymbol{footnote}}
%{\Huge \bf Accelerating, charged and rotating \\
%\vskip 5mm
%  NUT black holes - o -}
\vskip 9mm  
{\Huge \bf Rotating and swirling binary black
\vskip 4mm
   hole system balanced by its 
    \vskip 7mm 
  gravitational spin-spin interaction  }
\vskip 27mm
{\large {Marco Astorino$^{a}$\footnote{marco.astorino@gmail.com} ,
Matilde Torresan$^{b}$\footnote{matilde.torresan@studenti.unimi.it}
}}\\

\renewcommand{\thefootnote}{\arabic{footnote}}
\setcounter{footnote}{0}
\vskip 8mm
\vspace{0.2 cm}
{\small \textit{$^{a}$Laboratorio Italiano di Fisica Teorica (LIFT),  \\
Via Archimede 20, I-20129 Milano, Italy}\\
} \vspace{0.2 cm}
%{\small \textit{$^{b}$Istituto Nazionale di Fisica Nucleare (INFN), Sezione di Milano \\
%Via Celoria 16, I-20133 Milano, Italy}\\
%} 
%\vspace{0.2 cm}

{\small \textit{$^{b}$Universit\`a degli Studi di Milano}} \\
{\small {\it Via Celoria 16, I-20133 Milano, Italy}\\}

\end{center}

\vspace{2 cm}

\begin{center}
{\bf Abstract}
\end{center}
{We present the first exact and analytical solution in General Relativity describing an equilibrium configuration for two stationary black holes. The metric models two collinear  extremal Kerr black holes immersed in an external and back-reacting rotating tidal drag. The gravitational attraction is balanced by the repulsive gravitational spin-spin interaction generated by the interplay between black holes angular momenta and the rotational background. The new solution is built by embedding the double Kerr metric into a swirling universe by means of the Ehlers transformation. The geometry is completely regular outside the event horizons. Thermodynamic properties of the binary black hole system are studied, the Smarr law, the first law and the Christodoulou-Ruffini formulas are verified.
%From the analysis of the specific heat we show that the rotating binary can be thought of as the final state of an unstable single black hole elongated and stripped by the tidal forces of the rotating background. 
Microscopic degrees of freedom of the entropy are computed from the dual CFT living on the boundary of the near horizon geometries.}

\end{titlepage}

\addtocounter{page}{1}

\newpage

%\tableofcontents
%\newpage

\section{Introduction}
\label{sec:introduction}

The discovery of the double rotating black hole of Kerr type in 1980 \cite{KramerNeug}, opened to a possible equilibrium configuration between spinning black holes. In fact, in general relativity, a repulsive non-Newtonian effect, involving the angular momenta of the masses can occur\footnote{Actually repulsive interactions in gravity are not uncommon. Even in the Newtonian solar system model, planets, due to their revolutionary motion, follow stable elliptical orbits and they do not collapse on the Sun.}. Unfortunately it has been proven that it is not possible to balance the gravitational attraction of the two Kerr sources by means of their gravitational spin-spin interaction \cite{wald}, \cite{Herdeiro:2008kq}, because the angular momentum of at least one of the black holes must increase to the point of reaching a hyper-extremal configuration \cite{dietz}. That means that at least one of the two sources is a naked singularity. 

While this is the case for a couple of standard Kerr masses in an asymptotically flat scenario, things may change in the presence of an extra rotational background whose tidal forces affect the angular momentum of the black hole and non-linearly interact with it, without breaking the event horizon. To do so, we plan to embed the asymptotically flat double Kerr solution into the rotational background provided by the so called {\it ``swirling universe''} \cite{swirling}; this is a regular axially symmetric stationary rotating spacetime with a gradient of rotation growing in the azimuthal direction, so it spins in an opposite way for different positivity of the $z$-coordinate and it is null on the equatorial plane\footnote{The swirling background metric was probably firstly found in \cite{harrison}, even though it was not physically interpreted there. More properties about black holes in the swirling background can be found in \cite{swirling} and \cite{Moreira:2024sjq}.}. We can think of it as the rotational equivalent of the Bonnor-Melvin magnetic universe; in fact it shares with it numerous properties, but it does not need an electromagnetic field strength (nor any other energy momentum tensor). First of all, both spacetimes are not asymptotically flat. Then as the magnetic universe can be analytically  generated by a Lie-point symmetry of the Ernst's equations, the Harrison transformation \cite{ernst-magnetic}, the swirling background can be obtained thanks to the Ehlers transformation of the said equations. Actually, because of this symmetry transformation, we are able to immerse any given axisymmetric and stationary spacetime into the swirling universe \cite{swirling}. This strategy is inspired by a recent result \cite{marcoa-remove} about the removal of the conical singularity from the rotating C-metric, precisely because of the interaction of the black hole's angular momentum with the swirling background. Since one can consider the C-metric to be a near horizon close-up to the biggest constituent of a binary black hole system, this procedure may reveal successful outside the near horizon limit of a black hole couple too.

Thus, the aim of this paper is to construct the new solution as a one-parameter generalization of the double Kerr spacetime, where the additional parameter -  introduced via a Lie-point symmetry - corresponds to the strength of the external  back-reacting  rotational tidal forces. This will be done in section \ref{sec:generation}. Then, in section \ref{sec:equilibrium}, the axis of symmetry is studied to understand if equilibrium configurations of the two spinning black holes are possible; that would mean removing all the possible conical singularities from the spacetime. In section \ref{sec:thermo} some physical and thermodynamic properties of the new spacetime are examined, such as the mass, the angular momentum, the Smarr law and the second law of thermodynamics. Finally, in section \ref{sec:cft}, the duality of the near horizon geometry of the extremal event horizons and a two dimensional conformal field theory allows us to compute the entropy of the system from a microscopic perspective. Moreover the relation between the near horizon geometry of the swirling binary and the standard extremal Kerr metric helps us confirm previous results on the conserved charges. 

In the realm of general relativity,  exact, analytical and regular (outside the event horizons) black hole binary solutions at equilibrium are very scarce. According to the authors' knowledge the only completely regular one is given by a binary system inside an expanding bubble of nothing \cite{bubble}. There are also other examples of Bach-Weyl binaries regularised by a multipolar external gravitational field; for example, consider \cite{marcoa-binary}, \cite{many-rotating} and \cite{multipolar-acc} for charged, rotating and accelerating systems. However these latter examples can be better considered as local models not too far from the sources, because the curvature scalar invariants grow unbounded at spatial infinity close to the azimuthal axis. In the presence of the Maxwell electromagnetic field the spectrum of possibilities is wider because the electromagnetic force can balance the gravitational attraction. In fact, some well-known solutions are the Majumdar-Papapetrou spacetime \cite{majumdar}, \cite{papapetrou}, \cite{hartle-hawking}, its cosmological generalisations \cite{Kastor-MP}, \cite{Klemm-MP} or the black di-hole \cite{emparan-dihole}, \cite{emparan-teo}. Few other examples have been recently obtained numerically in the presence of a cosmological constant \cite{Dias:2023rde},\cite{Dias:2024dxg} or of a complex scalar field \cite{Herdeiro:2023mpt},\cite{Herdeiro:2023roz}. 

Nevertheless in Einstein's theory of gravitation a rotating analytical exact solution representing a binary black hole, regular outside the horizon, is still missing. The purpose of this article is to fill this gap, while providing an unexpected way to balance the equilibrium configuration.\\

\section{Double Kerr black holes in the swirling universe}
\label{sec:2KerrSwirling}

\subsection{Generation of the extremal solution}
\label{sec:generation}

As discovered by Ernst in \cite{ernst1}, Einstein's field equations for pure general relativity in vacuum, i.e. $ R_{\m\n}=0, $
can be cast, for axisymmetric and stationary spacetimes described by the Lewis-Weyl-Papapetrou metric,
\beq \label{lwpm}
     {ds}^2 = -f ( d\varphi - \omega dt)^2 + f^{-1} \bigl[ \rho^2 dt^2 - e^{2\gamma}  \bigl( {d \rho}^2 + {d z}^2 \bigr) \bigr] \ ,
\eeq 
as the complex Ernst equation
\beq \label{Ernst-eq}
       \bigl( \Er + \Er^* \bigr) \nabla^2 \Er   =   2 \, \vec{\nabla} \Er  \cdot \vec{\nabla} \Er  \ ,
\eeq
where the gravitational Ernst potential $\Er(\rho,z)$ is defined as 
\beq \label{Ernst-pot}
             \Er \coloneqq f + i h   \ , \qquad \qquad 
\eeq
such that
\beq
\label{h}
\vec{\nabla}   h \coloneqq - \frac{f^2}{\rho} \vec{e}_\varphi \times \vec{\nabla} \omega \ \ .
\eeq
The differential operators are thought as standard flat cylindrical Laplacian and gradient in three spatial dimensions (spanned by the orthonormal base $\{ \vec{e}_\rho, \vec{e}_z, \vec{e}_\varphi\}$). Because of the stationarity and axisymmetry, the spacetime geometry possesses at least a commuting couple of independent Killing vectors ($\p_t, \p_\varphi$). Furthermore all the functions depend, at most,  on the non-Killing coordinates, the Weyl ones ($\rho,z$).  \\
The advantage of Ernst's equation (\ref{Ernst-eq}), with respect to the equivalent vacuum Einstein equations $R_{\m\n}=0$, consists in the fact that the relevant degrees of freedom are properly separated, so the field equations for determining the $\g(\rho,z)$ function in (\ref{lwpm}) are completely decoupled from the system. Hence the $\g(\r,z)$ function can be determined after obtaining $f(\rho,z)$ and $\om(\rho,z)$. Another advantage of the  Ernst's equation is that it is explicitly invariant under the Ehlers transformation   
\beq
\label{ehlers-transf}
  \Er \to \hat{\Er} = \frac{\Er}{1+i\jmath\Er} \ .
\eeq
The above transformation maps a given axisymmetric and stationary solution of the Einstein's equations, which is called {\it seed}, into another vacuum solution of the same kind. $\jmath$ is the parameter labelling the Lie-point transformation, which characterises the rotational intensity of the background gravitational field. In this case note that the $\g$ function remains invariant under the Ehlers transformation; therefore there is no need to explicitly write down the system of partial differential equations $\g$ obeys, for more details about that see \cite{enhanced}. Note however that the $\g$ field equations are not included in (\ref{Ernst-eq}), which instead are a couple of real equations for determining $f$ and $\om$.\\
The Ehlers transformation acts on a given seed in two different ways: depending on whether it is combined with the discrete conjugation symmetry\footnote{The conjugation symmetry relates the two inequivalent versions of the Lewis-Weyl-Papapetrou metrics, the one in (\ref{lwpm}) and $\tilde{ds}^2 = -\tilde{f}(dt-\tilde{\omega}d\varphi)^2 +\tilde{f}^{-1}[e^{2\tilde{\gamma}}(d\r^2+dz^2) + \r^2d\varphi^2]$. In practice, it consists in the following discrete transformation: $\{\tilde{f} \to -\r^2/f + f\omega^2, \ \tilde{\omega} \to f^2\omega/(f^2\omega^2-\r^2), \ e^{2\tilde{\g}} \to e^{2\g} (\r^2/f^2-\omega^2)\}$; note that this transformation is not a Lie-point symmetry, meaning that it does not introduce any continuous parameter associated to an integrating constant. When the Ehlers transformation (\ref{ehlers-transf}) is applied with the Lewis-Weyl-Papapetrou metric conjugate to (\ref{lwpm}), instead of adding an external rotational background, it adds the NUT parameter to the seed metric.} of the Lewis-Weyl-Papapetrou metric, as described in \cite{swirling}, it could add\footnote{The actual effect of this kind of Ehlers transformation is the partial or total rotation of the mass into the NUT parameter. Its action on an asymptotically flat spacetime is studied \cite{PD-NUTs}.} the NUT parameter \cite{enhanced} or it could embed the seed into a rotating background dubbed swirling universe \cite{swirling}. Often in the first case it is called $electric$, while in the second $magnetic$ Ehlers transformation.

As a preliminary exercise, we apply the (magnetic) Ehlers transformation (\ref{ehlers-transf}) to the Minkowski spacetime to generate a new regular solution of the vacuum Einstein equations representing a rotating universe, called the {\it swirling universe}. To do so, we consider as seed the Minkowski metric in cylindrical coordinates 
\beq \label{minkowski}
        ds^2 = - dt^2 + d\rho^2 + dz^2 + \rho^2 d\varphi^2 \ .
\eeq 
The above metric (\ref{minkowski}) can be written in terms of the Lewis-Weyl-Papapetrou one, as in (\ref{lwpm}), just by setting $f=-\r^2, \ \omega =0, \ \g = \log \r$. Then, thanks to the definitions (\ref{Ernst-pot}) and (\ref{h}) we can write the Ernst potentials for the Minkowski spacetime
\beq
         \Er = - \r^2 \ .
\eeq
When we apply the Ehlers transformation (\ref{ehlers-transf}) to the above Ernst potential, we get a new Ernst potential representing another solution of Einstein vacuum field equations, i.e.
\beq
         \hat{\Er} = \frac{-\r^2}{1 - i \jmath \r^2} \ .
\eeq
To write this solution in metric form, one has only to make use of the definitions (\ref{Ernst-pot}) and (\ref{h}) to get the transformed metric functions 
\beq
        \hat{f} = - \frac{\r^2}{1+\jmath^2\r^4} \ , \hspace{2cm} \hat{\omega} = 4 \jmath z + \omega_0 \ .
\eeq
Since the $\g$ function is not affected by the Lie-point transformation the final form of the rotating background is
\beq
       \hat{ds}^2 = \frac{\r^2}{1+\jmath^2\r^4} \left[d\varphi-(4 \jmath z + \omega_0)\right]^2 + (1+\jmath^2\r^4) \left[-dt^2+dz^2+d\r^2\right] \ .
\eeq
This regular metric describes a stationary rotating spacetime, where the rotation around the axis of symmetry increases for growing values of the $z$ coordinate and of the transformation parameter $\jmath$. So the spacetime swirls in opposite directions on the two semi-planes defined by $z=0$, and there is no rotation for zero values of $z$ (when $\omega_0=0$, which can be always achieved by a trivial linear transformation between the Killing coordinates ($t,\varphi$)).

In this article we are interested in this latter case: we would like to generate, through the Ehlers map a binary rotating black hole solution that can balance its gravitational attraction thanks to the gravitational spin-spin interaction. The spin-spin interaction is a non-Newtonian effect peculiar of metric theories of gravity, such as general relativity, which can have a repulsive effect on the rotating masses of the system \cite{wald}. Unfortunately this effect is not sufficient for balancing a couple of standard isolated Kerr black holes, because the angular momentum of at least one of the two constituents has to be so big that it prevents the presence of the event horizon: at least one of the two sources has to be hyper extremal \cite{dietz}, \cite{Herdeiro:2008kq}. 

Therefore our plan to take advantage of the spin-spin interaction, without spoiling the black hole interpretation, consists in adding an external, back reacting rotation, as the one introduced by the Ehlers map (\ref{ehlers-transf}). The rotating background should interact with the  black binary influencing the angular momentum of the components and providing the necessary repulsion to balance them. \\
The simpler setting one might think of is the embedding of the double Schwarzschild black hole, i.e. the Bach-Weyl metric \cite{bach-weyl}, into the swirling universe. Unfortunately, this approach is not effective because the black holes do not acquire angular momentum despite the rotating tidal forces; therefore the spin-spin interaction is absent. This attempt is summarized in appendix \ref{app:bach-weyl-swirling}. So it seems that the seed angular momentum is necessary in our scheme; then the next simplest black hole binary solution we can consider is the double Kerr black hole. In order to keep the computation as simple as possible we focus on a specialization of the Kerr binary where the two sources have the same mass but opposite angular momenta and moreover they are extremal \cite{manko}. The counterrotating feature is relevant to keep the generated solution simple because the rotating background has opposite rotation with respect to the equatorial plane too, while the extremality and the mass symmetry reduce the complexity of the spacetime to only two integration constants.  After the change to prolate spherical coordinates $(x,y)$, defined as
\beq \label{weyl-prolate}
           \r = k \sqrt{(x^2-1)(1-y^2)}  \ , \hspace{1.8cm}   z = k x y \ , 
\eeq
the generic  axisymmetric and stationary metric can be written in terms of functions of the Lewis-Weyl-Papapetrou metric (\ref{lwpm}) as\footnote{Note that in \cite{manko} the functions $f, \om, \g$ look different just because here we are considering the conjugate version of the Lewis-Weyl-Papapetrou metric, more details about the conjugate LWP metric can be found in \cite{swirling}. However the two metrics are exactly the same. If we used the LWP metric without conjugation we would obtain after the Ehlers transformation an extremal double Kerr-NUT binary.}
\beq \label{lwpmxy}
       ds =  - f (d \varphi - \omega dt)^2  + \frac{1}{f} \left[ k^2 (x^2-1)(1-y^2) dt^2 - e^{2\gamma} k^2 (x^2-y^2) \left( \frac{{d x}^2}{x^2-1} + \frac{{d y}^2}{1-y^2} \right) \right] \ ,
\eeq
with 
\bea \label{seed-inizio}
        f(x,y) &=& -\frac{D}{N} \r^2 + \frac{16 W^2 y^2 \a^2 \r^4}{k^2 D N}  \ , \\
        \om(x,y)&=& - \frac{4 k N W y \a}{D^2 k^2-16 W^2 y^2 \a^2 \r^2}  \ , \nn \\
        \g(x,y) &=&  \frac{D^2 k^2 \r^2 -16 W^2 y^2 \a^2 \r^4}{k^2 N (x^2-y^2)^4 (1+\a^2)^2}  \nn \ ,
\eea 
where
\bea
        N(x,y) &=&  [(x^2-1)^2 + \a^2 (x^2-y^2)^2]^2 - 16 \a^2 x^2 y^2 (x^2-1)(1-y^2) \ , \nn \\
        D(x,y) &=&  \{ x^4-1+\a^2(x^2-y^2)^2 + 2x[b(x^2-1)+c\a(x^2-y^2)] \}^2 + 4 [\a(x^2+y^2-2x^2y^2)+cx(1-y^2)]^2  , \nn \\
        W(x,y) &=& c \a (x^2-y^2)(3x^2+y^2) + b (3x^4+6x^2-1) + 8x^3 \ , \nn \\ 
        k &=& M \frac{a^2+M^2}{a^2-M^2} \ , \hspace{1cm}  \a \ = \ \frac{2Ma}{a^2-M^2}\ , \hspace{1cm}  b \ = \ \frac{a^2-M^2}{a^2+M^2}\ , \hspace{1cm}  c \ = \ \frac{2Ma}{a^2+M^2} \ .
\label{seed-fine}
\eea
The two integration constants of the solution are $a$ and $M$, where $a^2>M^2$ and $M>0$. They are roughly related to the angular momentum and the mass of the black holes, but also to the other relevant physical quantity which is the (coordinate) distance $d$ between the two black holes\footnote{In these coordinates the extremal black hole event horizons reduce to a point, so the distance between the black holes or their centres coincides.}, with
\beq \label{d}
          d = 2k =\frac{2M(a^2+M^2)}{a^2-M^2} \ .
\eeq
The positions of the two event horizons are located in $(x=1, \ y=\pm1)$. 
The condition $a^2>M^2$ does not mean that the solution is hyper-extremal, as would be in the single Kerr case \cite{Herdeiro:2008kq}, \cite{manko}. From the asymptotic falloff of the metric or of the conjugate\footnote{The Ernst potential conjugate to the one in (\ref{Ernst-pot}) is defined in a similar way but starting from the conjugate LWP metric with respect to the one in eq. (\ref{lwpm}), for more details see footnote 3 or \cite{swirling}. The explicit expression for the conjugate Ernst potential for the seed metric is given in \cite{manko}.} Ernst potential 
\beq \label{barE}
        \bar{\Er} = 1  - \frac{4 M}{r} + \frac{8 M^2}{r^2} +  \mathcal{O} \left(\frac{1}{r^3} \right) \ ,
\eeq 
in spherical coordinates (for large values of the radial coordinate, $ r = kx , y = \cos\theta)$, we can directly read the mass and the angular momentum of the black hole binary. Indeed comparing (\ref{barE}) with the Ernst potential of a generic asymptotically flat solution with mass $\tilde{M}$ and angular momentum $\tilde{J}$,
\beq
             \tilde{\Er} = 1 - \frac{2(\tilde{M}-i\tilde{B})}{r} + \frac{(z_* + 2 i \tilde{J}) \cos \theta + const }{r^2} + \mathcal{O} \left(\frac{1}{r^3} \right) \ ,
\eeq
where $\tilde{M}, \tilde{B}, \tilde{J}$ are respectively the total mass, NUT parameter, and angular momentum of the system, we infer that the total mass of the seed solution is $\tilde{M}=2M$ while its total angular momentum is zero $\tilde{J}=0$. These values reflect the fact that the constituents have the same mass $M$ and opposite angular momenta, because they are specular and counterrotating; however for more precise details about the conserved charges see section \ref{sec:thermo}.

Notice that this spacetime does not admit a non-trivial equilibrium configuration between the two gravitational sources, because of the dominant gravitational attraction. From a physical point of view it means that the two masses have to be sustained by the presence of a couple of external pulling cosmic strings or kept apart by a rod in between the two constituents, on the z-axis. From a geometrical point of view these strings or the rod are interpreted as conical singularities: a deficit or excess angle around the symmetry axis on the three different regions delimited by the two event horizons. So strictly speaking this spacetime should not be considered a pure vacuum solution, but a delta-like energy momentum tensor modelling the strings or the strut should be introduced. In the case of the strut the energy conditions are not respected, while the strings are infinitely long. Hence the spacetime is not regular outside the event horizons, bringing in some issues related to the fundamental hypothesis of smoothness of the manifold and the black hole interpretation. Moreover there is no astrophysical observation, nor other phenomenological or experimental trace of such conical singularities, so far. These are our main motivations to improve this picture by introducing a physical mechanism that can remove the conical singularities to remain with a regular spacetime in the domain of outer communications and to provide a mechanical reason to prevent the gravitational collapse by balancing the two black holes. The solution generating technique based on the symmetries of the Ernst equations can help us generalise the extremal double Kerr metric and it will add the necessary features to address these issues.     

To embed the above seed (\ref{seed-inizio})-(\ref{seed-fine}) into the rotating background we have first to find the seed Ernst potential $\Er$ as defined in (\ref{Ernst-pot}). First of all we need to  determine $h$ using the equation (\ref{h}), but taking into account that in the new  $(x,y,\varphi)$ coordinates, defined in (\ref{weyl-prolate}), the gradient takes the form 
\beq \label{gradient}
         \vec{\nabla} h  =  \frac{\vec{e}_x}{k} \sqrt{\frac{x^2-1}{x^2-y^2}} \ \p_x h \ + \ \frac{\vec{e}_y}{k} \sqrt{\frac{1-y^2}{x^2-y^2}} \ \p_y h \ + \ \frac{\vec{e}_\varphi \ \p_\varphi h}{k \sqrt{(x^2-1)(1-y^2)}} \ .
\eeq
The resulting $h(x,y)$ is
\bea 
      h = 2\frac{M}{a} (a^2+M^2)\left[3x-x^3+  (1-x^2)^2 \frac{H}{K} \right] \ ,
\eea
with 
\bea
        H &=& a^{10} 4 a^4 M^4 (a-M) (a+M) \left(-x^4 \left(12 y^2+1\right)+4 x^2 \left(y^4+9 y^2-3\right)+2 x^6-16 y^6+26 y^4-12 y^2+1\right) \nn \\ 
        &+& 2 a^4 M^4 x \left(a^2+M^2\right) \left(-3 x^4 \left(8 y^2+3\right)+x^2 \left(30 y^4+36 y^2+1\right)+5 x^6+6 y^2 \left(-8 y^4+y^2+2\right)-9\right) \nn \\ 
        &+&2 a^2 M^2 \left(a^6-M^6\right) \left(12 \left(6 x^2-1\right) y^4-8 \left(3 x^4+7 x^2-1\right) y^2+6 x^6-7 x^4+28 x^2+1\right) \nn \\
         &+&a^2 M^2 x \left(a^6+M^6\right) \left(12 \left(7 x^2+3\right) y^4-8 \left(2 x^4+13 x^2+3\right) y^2+5 x^6-x^4+23 x^2+29\right) \nn \\
         &+&a^{10} (x-2) (x+1)^6+M^{10} (x-1)^6 (x+2) \ , \nn \\
        K &=&   8 a^4 M^4 x (a-M) (a+M) \left(-6 \left(x^2-3\right) y^4-6 \left(x^4-4 x^2+3\right) y^2+x^6+x^4-11 x^2-8 y^6+5\right)\nn \\ 
        &+& 2 a^4 M^4 \left(a^2+M^2\right) \left(-4 x^6 \left(6 y^2+1\right)+2 x^4 \left(6 y^2 \left(y^2+4\right)-7\right)+4 x^2 \left(-8 y^6+6 y^4-6 y^2+1\right) \right. \nn \\ 
        &+& \left. +5 x^8+8 y^8-4 y^4+1\right) + a^{10} (x-1)^2 (x+1)^6+M^{10} (x-1)^6 (x+1)^2 \nn \\ 
        &+&4 a^2 M^2 x \left(a^6-M^6\right) \left(4 \left(5 x^2-3\right) y^4-4 \left(3 x^4+4 x^2-3\right) y^2+3 x^6+x^4+9 x^2-5\right)\nn \\ 
        &+&a^2 M^2 \left(a^6+M^6\right) \left(-16 \left(x^4+6 x^2-3\right) x^2 y^2+8 \left(1-3 x^2\right)^2 y^4+5 x^8+4 x^6+38 x^4-12 x^2-3\right)\nn \ .
\eea
Then, thanks to (\ref{Ernst-pot}), we can write the seed Ernst potential $\Er=f+ih$ for the seed metric (\ref{seed-inizio}). This value can be plugged into the Ehlers transformation (\ref{ehlers-transf}) to get the new solution in terms of the Ernst potential
\beq
              \hat{\Er} = \frac{\Er}{1+i\jmath\Er} = \hat{f} + i \hat{h}
\eeq
To express this solution in the metric form one needs to extract $\hat{f}$ as the real part of $\hat{\Er}$: 
\beq \label{fhat}
             \hat{f}(x,y) = \frac{f}{(1-\jmath h)^2 + \jmath^2 f^2}
\eeq
and solve the eq. (\ref{h}) from the imaginary part of the transformed Ernst potential, i.e.  $\hat{h}$, to get $\hat{\om}(x,y)$, which is reported in appendix \ref{app:omega}. Finally the generated metric reads\footnote{A Mathematica notebook containing the generated solution is provided in the arxiv source files.} 
\beq \label{2-swriling-kerr}
       \hat{ds}^2 =  -\frac{f ( \D_\varphi d\varphi - \hat{\omega} dt)^2}{(1-\jmath h)^2 + \jmath^2 f^2}  + \frac{(1-\jmath h)^2 + \jmath^2 f^2}{f} \left[ k^2 (x^2-1)(1-y^2) dt^2 - e^{2\gamma} k^2 (x^2-y^2) \left( \frac{{d x}^2}{x^2-1} + \frac{{d y}^2}{1-y^2} \right) \right]  .
\eeq
This metric represents the extremal and counterrotating double Kerr black hole binary embedded into the swirling universe. $\jmath$ is the real parameter of the Lie-point transformation (\ref{ehlers-transf}), so when it vanishes the transformation becomes the identity map and the metric recovers the locally asymptotically flat seed (\ref{lwpm}), (\ref{seed-inizio})-(\ref{seed-fine}). $\jmath$ can be considered an extra integrating constant and it determines the intensity frame dragging due to the rotating background. This rotating background is thought to be generated by a couple of rotating black holes at infinity that are infinitely separated on the axis of symmetry, one for positive $z$ and the other for negative $z$. This conjecture is based on the fact that the Bonnor-Melvin universe is generated by a couple of Reissner-Nordstrom black holes infinitely far away \cite{Emparan:2001gm}. Moreover, both the swirling and Bonnor-Melvin universes can be generated by the double Wick rotation of an electric Lie-point transformation of the Schwarzschild black hole, the Ehlers and the Harrison transformations respectively. Thus, the double Wick rotation in this context plays the role of transforming the single black hole into a couple and pushes them away in opposite directions on the $z$-axis; so far away that the singularity disappears, but the rotational frame dragging or the electromagnetic field does not. \\
Note that, because of (\ref{fhat}), $g_{\varphi\varphi}$ shares the same sign of the seed $g_{\varphi\varphi}$ by construction; hence if the seed is not affected by closed time-like curves, the same property is inherited by the generated metric, after the Ehlers transformation.  \\
We have introduced the gauge parameter $\D_\varphi$ to better control the presence of conical singularities as described in the section below; it basically corresponds to a rescaling of the angle $\varphi$ or equivalently to a change of its range. We prefer to leave the range of the azimuthal angle as usual, i.e. $\varphi \in [0,2\pi)$, while showing explicitly in the metric $\D_\varphi$, which eventually can be fixed to enhance the metric smoothness. \\

\subsection{Removal of conical singularities: rotating black hole binary at equilibrium}
\label{sec:equilibrium}

\begin{figure}[htbp]
    \centering
    \begin{minipage}{0.45\textwidth}  % Adjust width as necessary
        \centering
        \includegraphics[scale=0.46]{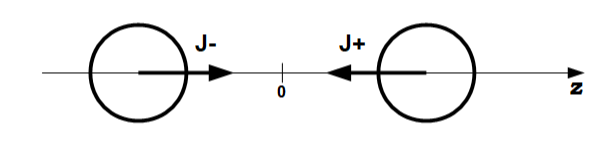}\par
        \captionsetup{
    font={scriptsize},         % small, italic text
    labelfont={bf},          % bold label (e.g., "Figure")
    justification=centering  % center the caption text
    }
    \caption{Counter-rotating configurations of two black holes with opposite angular momentum.}
        \label{fig:rep1}
    \end{minipage}\hfill
    \begin{minipage}{0.45\textwidth}
        \centering
        \includegraphics[scale=0.46]{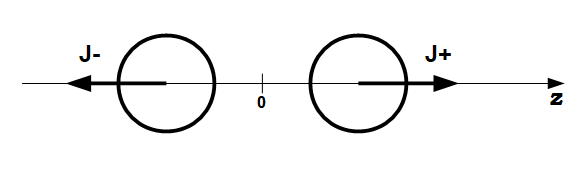}\par
        \centering
    \captionsetup{
    font={scriptsize},         % small, italic text
    labelfont={bf},          % bold label (e.g., "Figure")
    justification=centering  % center the caption text
    }
    \caption{Counter-rotating configurations of two black holes with opposite angular momentum.}
    \label{fig:rep2}
    \end{minipage}
\end{figure}

\begin{figure}[htbp]
    \centering
    \begin{minipage}{0.45\textwidth}  % Adjust width as necessary
     \hspace{3mm} 
       \centering
        \includegraphics[scale=0.46]{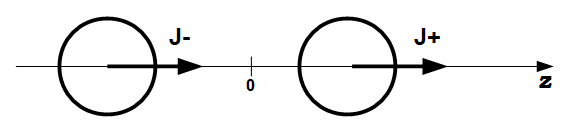}\par
        \captionsetup{
    font={scriptsize},         % small, italic text
    labelfont={bf},          % bold label (e.g., "Figure")
    justification=centering  % center the caption text
    }
    \caption{Co-rotating configurations of two black holes with aligned angular momentum.}
        \label{fig:att1}
    \end{minipage}\hfill
    \begin{minipage}{0.45\textwidth}
        \centering
        \includegraphics[scale=0.46]{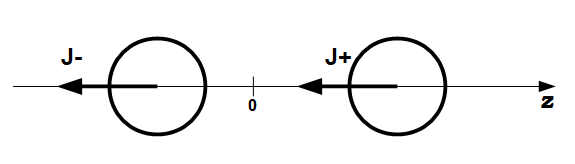}\par
        \centering
    \captionsetup{
    font={scriptsize},         % small, italic text
    labelfont={bf},          % bold label (e.g., "Figure")
    justification=centering  % center the caption text
    }
    \caption{Co-rotating configurations of two black holes with alined angular momentum.}
    \label{fig:att2}
    \end{minipage}
\end{figure}

In case we want to have a legitimate black hole binary configuration, the metric must not be plagued by conical singularities. To check if this is possible we compute on the $z$-axis, outside the event horizons, the ratio between a small circle around the z-axis $L=2\pi\sqrt{g_{\varphi\varphi}}$ and its polar radius $R= \r \sqrt{g_{zz}}$, that is
\beq
      \label{cony}      \lim_{\r \to 0} \frac{L}{R} \ = \ \lim_{\r \to 0} \frac{2\pi}{\r} \sqrt{\frac{g_{\varphi\varphi}}{g_{zz}}} \ .
\eeq
If this quantity equals $2\pi$ there are no conical singularities, otherwise we can have conical deficit or excess respectively when (\ref{cony}) is smaller or bigger than $2\pi$.
Note that to do this computation we are using the Weyl cylindrical coordinates $(t,\r,z,\varphi)$ related to the prolate spherical ones by the inverse transformation of (\ref{weyl-prolate}), which can be written as follows
\bea
          x &=& \frac{1}{2k} \left( \sqrt{\r^2+(z+k)^2} + \sqrt{\r^2+(z-k)^2} \right) \ , \nn \\ 
          y &=& \frac{1}{2k} \left( \sqrt{\r^2+(z+k)^2} - \sqrt{\r^2+(z-k)^2} \right) \ .
\eea
In these coordinates the black hole horizons are, because of the extremality, point-like and located at $ \r = 0 $ and $ z = \pm k $. So, in order to remain with a singular free axis of symmetry, the possible  conicity described in equation (\ref{cony}) has to be simultaneously normalized to $2\pi$ on the event horizon poles for all the three axis regions: for $z<-k$, $-k<z<k$ and $z>k$. These three conditions give us two constraints for the parameters of the solution
\beq \label{j-Dphi}
       ^{({1 \above -4pt 2})} \bar{\jmath} = \frac{\pm a (a\pm M )^2}{4 M (\mp a^4 + 2 a^3  M + M^3 \pm 2 a M^4)} \ , \hspace{0.7cm}  ^{({1 \above -4pt 2})} \bar{\D}_\varphi =\frac{16 a^6 M^2 }{(\mp a^4 + 2 a^3  M + M^3 \pm 2 a M^4)^2}
\eeq
Note that the couple ($\jmath, \D_\varphi$) can assume two different values which ensure the equilibrium configuration for the black holes, without string or struts. Henceforward the barred $^{({1 \above -4pt 2})}\bar{\jmath}$ and $^{({1 \above -4pt 2})}\bar{\D}_{\varphi}$ symbols will refer to the specific values in eq. (\ref{j-Dphi}), which regularise the spacetime. Actually only one physical parameter of the new solution has to be constrained, because $\D_\varphi$ is just a gauge parameter. The spacetime basically remains with two physical integration constants $a$ and $M$, as the seed metric, but thanks to the constraints (\ref{j-Dphi}) now it has been regularised, in the domain of outer communication. \\
The possible equilibrium configurations are two, depending on which of the two different directions of the swirling rotation is taken into consideration with respect to the seed initial angular momentum (encoded in the positivity of the $\jmath$ parameter). When the direction of rotation of the background is aligned with the angular momentum of the system, it increases the seed angular momentum. On the other hand, when $\jmath$ has opposite positivity relative to $a$, the black hole system acquires a new angular momentum in the opposite direction with respect to the double Kerr seed. Because of antisymmetry there are no other non-trivial orientations: basically the two equilibrium configurations exhaust all the allowed antiparallel spins disposal. This can be also intuitively understood from Wald's spin-spin estimate \cite{wald} of the potential energy between two rotating massive bodies
\beq \label{spin-spin-wald}
      U = - G \frac{m_1 m_2}{r} - \frac{G}{r^3}\left[\vec{J_1} \cdot \vec{J_2} - 3 (\vec{J_1} \cdot \vec{e_r})(\vec{J_2} \cdot \vec{e_r}) \right]  \ ,
\eeq
where $\vec{r} =r \vec{e_r}$ is the distance between the two sources (which in this case is aligned on the $z$-axis $\vec{e_z}$), $m_i$ and $\vec{J}_i$ their masses and angular momenta. From the formula (\ref{spin-spin-wald}) it is clear that the maximum repulsive force is exerted when the angular momenta are coaxial and mutually parallel, as represented in figures \ref{fig:att1}-\ref{fig:att2}, while it becomes maximally attractive when the angular momenta are coaxial and anti-parallel, as pictured in figures \ref{fig:rep1}-\ref{fig:rep2}. Heuristically this explains why the condition $a^2>M^2$ is fulfilled at extremality both for the seed and for the swirling binary. In fact, by construction,  we are considering the antiparallel orientation of the spin-spin interaction, which is the configuration that lowers the total energy of the system.

%\begin{figure}[h!]%
%\captionsetup[subfigure]{labelformat=empty}
%\centering
%\hspace{-0.8cm}
%\subfloat[\centering $m=1$, $c=0.05$]{{\includegraphics[scale=0.4]{spin-prova.png}}}%
%\hspace{1cm}
%\subfloat[\centering $m=1$, $c=0.1$]{{ \includegraphics[scale=0.4]{spin-prova.png}}} \\
%\subfloat[\centering $m=1$, $c=0.05$]{{\includegraphics[scale=0.4]{spin-prova.png}}}%
%\hspace{1cm}
%\subfloat[\centering $m=1$, $c=0.1$]{{ \includegraphics[scale=0.4]{spin-prova.png}}}%
%\caption{\small Fare le figure con tuttle le 4 configurazioni possibili ...}%
%\label{fig:spins}
%\end{figure}

So the repulsive interaction able to balance the binary system is not in line with (\ref{spin-spin-wald}) estimate, as also suggested by \cite{Herdeiro:2008kq}, probably because eq (\ref{spin-spin-wald}) comes from an approximation and do not take into consideration the full general relativistic effects.

From the eq. (\ref{cony}), it follows that when $\jmath=0$ the seed equilibrium configuration can be reached only for $\D_\varphi=1$ and $M^2=a^2$. Physically it means that they cannot experience their mutual gravitational attraction because the (coordinate) distance between the black holes becomes infinite. In this case the two constituents become two standard extremal black holes that are non-interacting because they are infinitely separated. \\
However, when $\jmath \neq 0$ equilibrium configurations are possible for finite separation between the two black holes, in terms of coordinate distance. On the other hand, the proper distance between the extreme binary constituents, along the axis of symmetry,
\bea
        d_p &=& \lim_{\r \to 0} \int_{-k}^k \sqrt{g_{zz}(\r,z)} \ dz   \ ,
       % &=&\int_{-k}^k 4 \sqrt{\frac{M^4(a-4a^2jM-4jM^3)^2[a^2M^2(a^2+M^2)^3(4a^4-3a^2M^2+M^4)+2M^2(a^3-aM^2)^2(-3a^4-2a^2M^2+M^4)z^2+a^2(a^2-M^2)^4z^4]}{(a^2+M^2)^4[M^2(a^2+M^2)^2-(a^2-M^2)^2 z^2]^2}}
\eea
diverges, because of the degeneracy of the inner and outer horizons. This is a property inherited from the seed, even if it is not at equilibrium. Indeed it is independent of our construction, but a feature of extreme horizons. In fact a notable case where this also happens is the Majumdar-Papapetrou charged binary at equilibrium, as explicitly shown in appendix \ref{app:MP}.\\

\subsection{Charges, Smarr and first law of Thermodynamics}
\label{sec:thermo}

\begin{figure}[h]
%	\captionsetup[subfigure]{labelformat=empty}
	\centering
%	\hspace{-0.2cm}
	\includegraphics[scale=0.33]{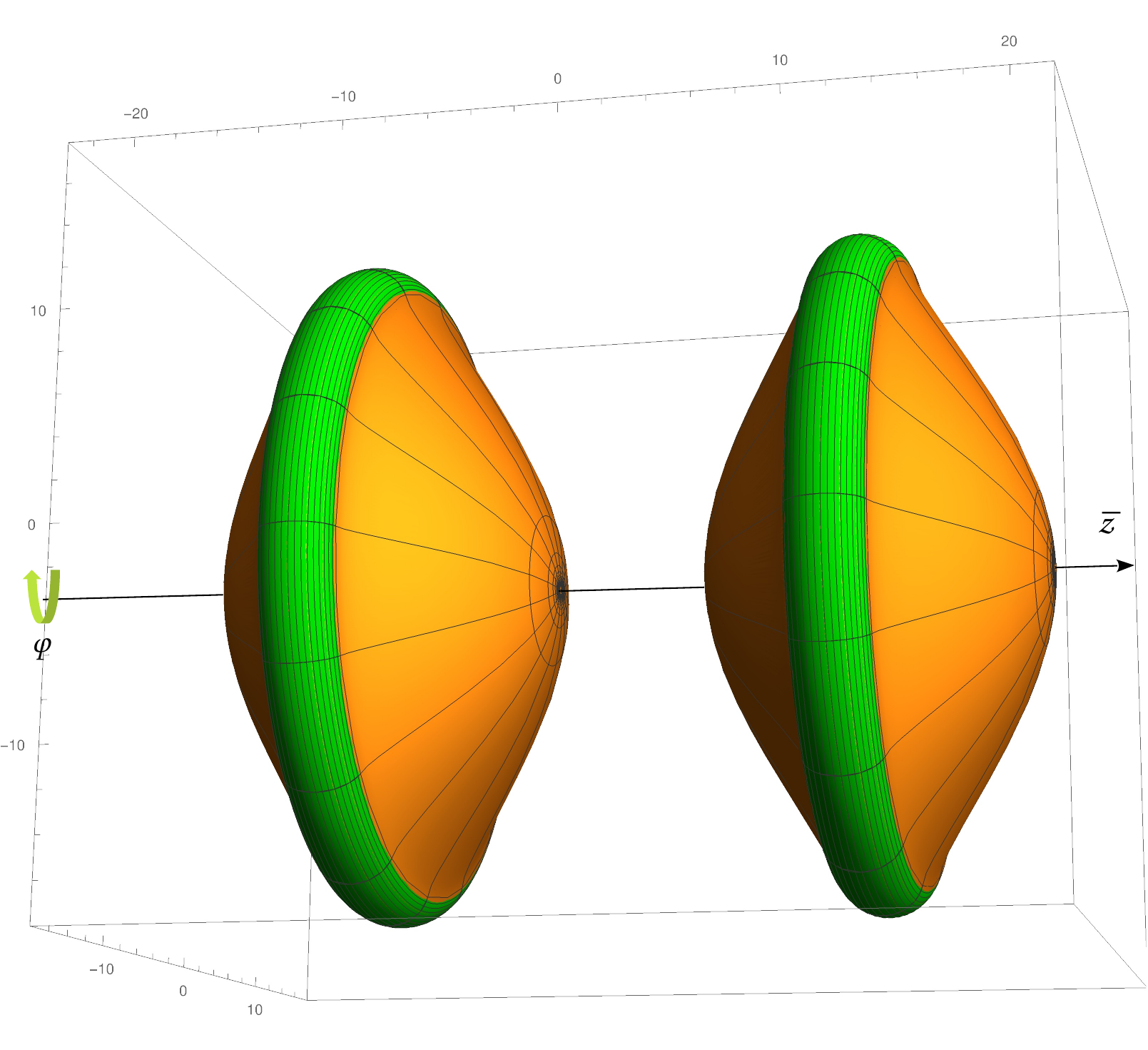}
	\caption{\small The event horizons of a regularised extremal counterrotating Kerr black binary immersed in the swirling universe, are isometrically embedded into the three-dimensional flat space, for $M=1, a=2$. Since extremal Kerr black hole cannot be globally embedded into the 3D Euclidean flat space \cite{smarr-embedding}, here we picture in green the portions of the horizons that can be embedded into the Euclidean space $ d\bar{\rho}^2 + \bar{\rho}^2 d\varphi^2 + d\bar{z}^2 $, while in orange the portions of the event horizons embedded into pseudo-euclidean space $ d\bar{\rho}^2 + \bar{\rho}^2 d\varphi^2 - d\bar{z}^2 $. Thanks to the regularisating constraints [$\jmath = ^{(1)}\bar{\jmath}, \D_\varphi = ^{(1)}\bar{\D}_\varphi$] of (\ref{j-Dphi}) the surfaces of the horizons are everywhere smooth. Similar images can be obtained with the other regularisation.}
	\label{fig:horizons}
\end{figure}

Computing physical properties of this extreme binary is non-trivial because, both in Weyl or in prolate spherical coordinates, the event horizons degenerate into a line, i.e. $[\r = 0 , \ z = \pm k, \ \varphi \in (0,2,\pi)]$ or $[x=1 , \ y=\pm 1, \ \varphi \in (0,2,\pi)]$ respectively, instead of a surface. Hence it is not possible to perform the integration over the event horizons in these coordinates. It is, thus, useful to switch to the coordinates of \cite{kodama}
\beq
        x = \sqrt{\frac{X^2+1}{X^2+Y^2}} \ , \hspace{1.4cm}  y = Y \sqrt{\frac{X^2+1}{X^2+Y^2}} \ , 
\eeq
which have the advantage of expanding the event horizon to a proper surface: $[X \geq 0, \ Y = \pm 1 , \ \varphi \in (0, 2\pi) ]$. Therefore these coordinates are particularly useful to picture the shape of the event horizons as isometrically embedded into a flat three-dimensional space. As can be seen in figure \ref{fig:horizons} the black hole geometries are substantially deformed from the high rotation needed to sustain the system at equilibrium. 

Moreover thanks to these coordinates we can write the Komar mass with respect to the timelike Killing vector field $\x=\p_t$
\beq
          m = -\frac{1}{8 \pi} \int_\S \e_{\m\n\l\s} \nabla^\m \xi^\n dx^\l \w dx^\s  \ ,
\eeq
where the orientation of the spatial two-surface of integration $\S$ must be chosen to remain with a positive orientation of the spacetime. In particular we have to take care that in $Y=\pm1$ the normal to the surface of the two horizons is opposite. So the masses for the two black holes, located at $Y = \pm 1$, are 
\bea \label{m}
      m_\pm &=& \mp \frac{1}{8\pi} \int_0^{2\pi} d\varphi \int_0^\infty dX \sqrt{|g|}  g^{YY} \left( g^{tt} \p_Y g_{tt} + g^{t\varphi} \p_Y g_{t\varphi} \right) \\
            &=&  \frac{M(a^3+4a^4 \jmath M -aM^2 + 8a^2 j M^3 +4 \jmath M^5)\D_\varphi}{(a^2-M^2)(a+4a^2\jmath M -4 \jmath M^3)}  .
\eea
Similarly the Komar conserved charge for the Killing vector $\z=\p_\varphi$ is given by 
\beq
      J = \frac{1}{16\pi} \int_\S \e_{\m\n\l\s} \nabla^\m \z^\n dx^\l \w dx^\s  \ .
\eeq
Then the angular momenta for the two black holes are
\bea \label{J}
      J_\pm &=&  \pm \frac{1}{16\pi} \int_0^{2\pi} d\varphi \int_0^\infty dX \sqrt{|g|}  g^{YY} \left( g^{tt} \p_Y g_{t\varphi} + g^{t\varphi} \p_Y g_{\varphi\varphi} \right) \\
            &=&  \pm \frac{a^3 M \D_\varphi^2}{-a^2 + 16 a^4\jmath^2M^2+8a\jmath M^3 -16 \jmath^2 M^6} \ .
\eea
So the masses for the two constituent are identical while the angular momenta are opposite.
In the limit $\jmath \to 0$ we recover the conserved charges of the seed black holes
\beq \label{mJ0}
            m_\pm\Big|_{\jmath=0} \ = \  M \D_\varphi \ \ , \hspace{1.5cm}  J_\pm\Big|_{\jmath=0} \ = \ \mp a M \D_\varphi^2 \ .
\eeq            
These values are coherent with the radial asymptotic falloff of the Ernst potential, as seen in section \ref{sec:generation}. Indeed, in that case, the equation (\ref{cony}) gives $\D_\varphi=1$ to avoid conicities, on the symmetry axis in the region external to the two black holes, which is the gauge choice of the seed\footnote{Alternatively for $\jmath=0$ one can remove the conical defect of the $z$-axis, in the region between the two black holes, by fixing $\D_\varphi=4a^2M^2/(a^2+M^2)^2$.}. \\
The angular velocity of the horizons is given by 
\beq
         \Om_\pm = - \frac{g_{t\varphi}}{g_{\varphi\varphi}}   \Big|_{Y=\pm1} =  \pm \ \frac{(-a+4a^2\jmath M + 4 \jmath M^3)(a^3+4a^4\jmath M -a M^2 + 8 a^2 \jmath M^3 + 4 \jmath M^5)}{2 a^3 (a^2-M^2) \D_\varphi} \ ,        
\eeq
while it is zero on the equatorial plane, because of the $\Z_2^{\varphi}$ symmetry\footnote{We refer to $\Z_2^{\varphi}$ symmetry, as done in \cite{Herdeiro:2008kq}, with the specular symmetry encoded in the invariance under the inversion transformation  $(z \to -z, \varphi \to - \varphi)$. In practice it means that the black holes have the same mass but their angular momenta are opposite, as can be appreciated from eqs. (\ref{m-J}).} across the $z=0$ plane. The black holes' temperature is also zero because they are extremal. To check this, it is sufficient, following \cite{tomimatsu-84}, to remove the conical singularity from the euclidean $(t,X)$ sector of the metric or to compute the surface gravity on the event horizons, which in $(t,X,Y,\varphi)$ coordinates is\footnote{The formula of \cite{tomimatsu-84} looks different only because refers to the conjugate Lewis-Weyl-Papapetrou metric with respect the one here used, (\ref{lwpm}) or (\ref{lwpmxy}). For more detail about conjugate LWP see \cite{swirling}.}
\beq
          T  \ = \ \lim_{Y \to \pm 1} \ \frac{1}{2\pi X } \ \sqrt{\frac{g_{tt}(g_{tt}g_{\varphi\varphi}-g^2_{t\varphi})}{g_{t\varphi}^2 g_{XX}}} \ = \ 0 \ .
\eeq
The area of the two black hole components is equal and given by
\beq \label{A}
       A \ = \ \lim_{Y \to \pm 1} \  \int_0^{2\pi} d\varphi \int_0^\infty dX \sqrt{g_{XX}g_{\varphi\varphi}} \ = \ \frac{16 \pi a^2 M^2}{a^2+M^2} \D_\varphi \ .
\eeq
The Bekenstein-Hawking entropy is taken as the quarter of the horizon area, that is $S = A/4$. A microscopic derivation of the black holes' entropies is given in section \ref{sec:cft} by a near horizon duality with a conformal field theory model. \\ 
Then it is easy to verify that the Smarr law 
\beq
         m_\pm = T  S + 2 \Omega_\pm J_\pm  
\eeq
holds individually for each element of the binary, thus for the binary system as a whole. 
On the other hand the first law of black hole thermodynamics does not hold in general for the above physical quantities. However we can define a specific observer renormalising the time coordinate by a constant factor $\D_t(M,a)$:
\beq
      t \to t \  \D_t \ ,
\eeq
such that the first law, 
\beq \label{first-law}
            \d \tilde{m}_\pm = \tilde{T} \d S + \tilde{\Om}_\pm \d J_\pm \ ,
\eeq
is fulfilled for each black hole, thus for the complete system. Note that the time rescaling factor $\D_t$ enters trivially both into the Komar mass and the angular velocity definitions, rescaling them as follows
\beq
     \tilde{T} = \D_t T  \ , \hspace{1.3 cm}   \tilde{m}_\pm = \D_t m_\pm \ , \hspace{1.3 cm} \tilde{\Om}_\pm = \D_t \Om_\pm \ .
\eeq 
A proper normalisation of the time coordinate is often required outside Minkowskian asymptotic, even for obtaining the proper energy of the Kerr-AdS solution. \\
Specifically we are interested in the analysis of the regularised model. In the presence of conical singularities, the first law differs from the standard one, additional terms might be taken into account. Moreover when the two regularising constraints  (\ref{j-Dphi}) are imposed, the relevant thermodynamic quantities simplify to
\bea
         ^{(1)}m_\pm &=& \frac{16 a^5 M^3}{(a-M)(M^2+2aM-a^2)^2 (a^2+M^2)} \ , \hspace{0.8cm}  ^{(2)}m_\pm \ = \ \frac{16 a^5 M^3}{(a+M)(a^2+2aM-M^2)^2 (a^2+M^2)} , \nn \\
          ^{(1)} \Om_\pm &=& \pm \frac{(a^2+M^2)^2}{4 a^3 M (a+M)} \ , \hspace{4.2cm} ^{(2)} \Om_\pm \ = \ \mp \frac{(a^2+M^2)^2}{4 a^3 M (a+M)} \ , \label{m-J} \\
           ^{(1)}J_\pm &=& \frac{\pm \ 32 a^8 M^4}{(a^2+M^2)^3(M^2+2aM-a^2)^2} \ , \hspace{2.1cm}  ^{(2)}J_\pm \ =  \frac{\mp \ 32 a^8 M^4}{(a^2+M^2)^3(a^2+2aM-M^2)^2} \ . \nn
\eea
But when the time normalising factors, one for each possible regularisation, 
\beq
          ^{(1)} \D_t = \frac{(a-M)|M^2+2aM-a^2|}{2\sqrt{2}aM\sqrt{a^2+M^2}} \ , \hspace{1.5cm}  ^{(2)} \D_t = \frac{(a+M)(a^2+2aM-M^2)}{2\sqrt{2}aM\sqrt{a^2+M^2}}
\eeq
are taken into consideration we obtain the following values for the black holes mass and angular velocity
\bea
         ^{(1)}\tilde{m}_\pm &=& \frac{4 \sqrt{2} a^4 M^2}{|M^2+2aM-a^2| (a^2+M^2)^{3/2}}   \ , \hspace{1.1cm} ^{(1)}\tilde{\Om}_\pm \ = \ \pm \frac{|M^2+2aM-a^2|(a^2+M^2)^{3/2}}{8 \sqrt{2} a^4 M^2}  \ , \nn \\
           ^{(2)}\tilde{m}_\pm &=&  \frac{4 \sqrt{2} a^4 M^2}{(a^2+2aM-M^2) (a^2+M^2)^{3/2}}  \ , \hspace{1.1cm} ^{(2)}\tilde{\Om}_\pm \ = \ \mp \frac{(a^2+2aM-M^2)(a^2+M^2)^{3/2}}{8 \sqrt{2} a^4 M^2}  \ . \qquad \label{m-tilde}
\eea 
These quantities, by construction, fulfil not only the Smarr law 
\beq
     ^{(i)} \tilde{m}_\pm =  2 \, ^{(i)}\tilde{\Omega}_\pm \, ^{(i)} J_\pm \ ,
\eeq
and the Christodoulou-Ruffini formula 
\beq
     ^{(i)} \tilde{m}_\pm^2 = \frac{S}{4\pi} + \frac{\pi \ ^{(i)}J_\pm^2}{S}   \ ,
\eeq
but also the first law of black hole thermodynamics, as stated in eq. (\ref{first-law}). Since $M^2=J$ we can infer some other relations between physical quantities, for instance
\beq
     S = 2 ( 1 \pm \sqrt{2} )\, J \ , \hspace{2cm} ^{(i)} \tilde{m}_\pm = \frac{1}{2 \, | ^{(i)} \tilde{\Omega}_\pm |}
\eeq

Note that $\D_t(M,a)$ is defined up to a numerical multiplying factor that can be fixed by requiring that the mass of the black holes has to be the mass of the single constituent, i.e. $M$, for infinite separation distance, when they do not suffer from mutual interaction.\\
The final values for the mass and angular momentum agree with the near horizon analysis of the next section.  \\

\subsection{Microscopic entropy from near horizon/CFT correspondence} 
\label{sec:cft}

The framework of the Kerr/CFT correspondence provides a tool to compute the entropy of a black hole from the analysis of its near horizon geometry \cite{strominger-kerr-cft}, \cite{strominger-duals}, \cite{compere-kerr-cft}. It relies on some universal thermodynamic properties of conformal field theories (CFT) in two dimensions, without the need of a detailed description of the theory. This scheme is particularly effective in the case of extreme objects and it has shown to work well also for multi-centred solutions \cite{enhanced}, asymptotically non-trivial background, such as accelerating \cite{c-cft}, swirling \cite{marcoa-remove} or electromagnetic Bonnor-Melvin \cite{magnetised-RN-cft}, \cite{magnetised-kerr-cft} universes. This method is based on a duality between the black hole microstates and the ones of a conformal field theory living on the asymptotic boundary of the near horizon metric. Following the above references we can find the near horizons metric of our composite system by defining a reference frame close to the event horizons by the coordinate transformation  
\beq\label{change_NH}
       t(\tau):= \frac{r_0}{\b} \tau   \ , \hspace{0.8cm} \rho(r,y):= \b r_0 r \sqrt{1-y^2} \ , \hspace{0.8cm} z(r,y):= \pm k + \b r_0 r y \ , \hspace{0.8cm}   \varphi(\tau,\phi):= \phi + \Om_\pm \frac{r_0}{\b} \tau \ ,
\eeq
where, as in the previous sections, the $+$ and $-$ signs refer to the component of the system with positive or negative z respectively. As expected, from \cite{lucietti-kunduri}, the near horizon metric (\ref{2-swriling-kerr}) can be described by the warped and twisted product of $AdS_2 \times S^2$, as follows
\beq \label{twisted}
          ds^2 = \G_\pm(y) \left[ - r^2 d\tau^2 +\frac{dr^2}{r^2} + \s^2(y) \frac{dy^2}{1-y^2} + \y^2(y) \Big( d\varphi^2 - \k_\pm r d\tau \Big)^2 \right] \ ,
\eeq
with
\bea
       \G_\pm(y) &=& \frac{M^2}{a^2+M^2} \Big\{ \Big[ 8a^7 \jmath M + 16 a^8 \jmath^2 M^2 -24a^5\jmath M^3 -40 a^3 \jmath M^5 -8a\jmath M^7 + 16 \jmath^2 M^{10} \nn \\     &+&a^2(a^4+M^4) (1+64 \jmath^2 M^4) + 6a^4 M^2 (1+16\jmath^2 M^4) \Big](1+y^2)  \label{Gamma-psi-inizio} \\
%       \G(y) &=& \frac{M^2}{a^2+M^2} \left\{ \left[8\jmath M (a^7+2a^8\jmath M-3a^5M^4-5a^3M^4-aM^6+2\jmath M^9) + 16 \jmath^2 M^{10} + a^2(a^4+M^4) (1+64 \jmath^2 M^4) \right. \right. , \nn \\
        &\pm&   2y\Big[(a^3+4a^4\jmath M+aM^2-4\jmath M^5)^2 -4a^2M^2(a-4a^2\jmath M-4\jmath M^3)^2 \Big] \Big\} \ , \nn  \\
        \sigma(y) &=& 1 \ , \hspace{2.9cm} \k_\pm \ = \ \pm \frac{2 a^2 M-32 a^4 \jmath^2 M^3 -16 a \jmath M^4 +32 \jmath^2 M^7}{a\D_\varphi(a^2+M^2)} \ , \nn \\
        \psi(y) &=& \frac{4 a^2 M^2 \sqrt{1-y^2}}{(a^2+M^2) \ \G_\pm(y)} \D_\varphi \ , \hspace{2.8cm} r_0 \ = \ \frac{2 a M}{\sqrt{a^2+M^2}} \ . \label{Gamma-psi-fine} % \\
%        \k &=& \frac{2 a^2 M-32 a^4 \jmath^2 M^3 -16 a \jmath M^4 +32 \jmath^2 M^7}{a\D_\varphi(a^2+M^2)} \ , \nn          
\eea

From this geometry we can deduce the central charge of the symmetry algebra related to the dual two-dimensional CFT located on the asymptotic boundary of the near horizon metric. Regarding the appropriate boundary conditions we refer to  \cite{strominger-duals}, \cite{compere-kerr-cft},\cite{magnetised-RN-cft}. For extremal black holes the only non-null central charge is the left one, given by the following integral 
\beq
         c_{L_\pm} = 3 \k_\pm \int_{-1}^{1} \frac{\G_\pm(y) \s(y) \psi(y)}{\sqrt{1-y^2}} dy = \ \pm \frac{48aM^3(a^2 - 16a^4\jmath^2 M^2 - 8a\jmath M^3  + 16 \jmath^2 M^6		)}{(a^2+M^2)^2} \ .
\eeq
Then from the Cardy formula we are able to count the microcanonical degrees of freedom of the left sector of the two dimensional boundary conformal field theory, i.e. the CFT entropy
\beq
            \mathcal{S}_{CFT} = \frac{\pi^2}{3} c_L T_L \ .
\eeq
$T_L$ is the Frolov-Thorne vacuum, a stationary generalisation of the Hartle-Hawking static vacuum. It is defined geometrically, as the near horizon limit of the surface gravity, and for rotating extremal black holes it becomes
\beq
      T_{L_\pm} = \frac{1}{2\pi \k_\pm}  \ .
\eeq
Thus the resulting entropy of the boundary CFT models dual to the black hole near horizon geometries is given by
\beq
        \mathcal{S}_{CFT_{\pm}} = \frac{\pi^2	}{3} c_{L_\pm} T_{L_\pm} =  \frac{4a^2M^2\pi \D_\varphi}{a^2+M^2} = \frac{A}{4}
\eeq
Therefore each component of the black binary fulfils the Bekenstein-Hawking law: its entropy corresponds to a quarter of the horizon surface, as in (\ref{A}). So this property holds, as well, for the composite system. \\
Note that this result is independent of the regularisation from the conical singularity. If one wants to restrict this result to spacetimes without cosmic strings or struts, one has to consider the two specific regularising couple of values for $\jmath$ and $\D_\varphi$, as in eq. (\ref{j-Dphi}). In these cases the expressions (\ref{Gamma-psi-inizio})-(\ref{Gamma-psi-fine}) simplify considerably; for instance the near horizon geometry is given, for the regularised case determined by ($^{(1)}\bar{\jmath},\ ^{(1)}\bar{\Delta}_\varphi$), they reduce to 
\bea \label{near-1}
        ^{(1)}\G_\pm(y) &=& \frac{32 a^8 M^4 (1+y^2)}{(a^2+M^2)^3(a^2 - 2aM - M^2)^2} \hspace{2 cm}\s(y)  = 1  \\
        \psi(y)  &=&  \frac{2 \sqrt{1-y^2}}{1+y^2}    \hspace{5.2cm} ^{(1)}\k_\pm = \mp 1  \ . \nn
\eea
On the other hand for the second possible regularizing values ($^{(2)}\bar{\jmath},\ ^{(2)}\bar{\Delta}_\varphi$) from (\ref{j-Dphi}), the near horizon geometry is defined by 
\bea \label{near-2}
        ^{(2)}\G_\pm(y) &=& \frac{32 a^8 M^4 (1+y^2)}{(a^2+M^2)^3(a^2 + 2aM - M^2)^2} \hspace{2 cm}\s(y)  = 1 \nn \\
        \psi(y)  &=&  \frac{2 \sqrt{1-y^2}}{1+y^2}    \hspace{5.2cm} ^{(2)}\k_\pm = \pm 1  \ .
\eea
The fact that the swirling parameter disappears from the extremal near horizon geometry means that the effect of the external rotation vanishes close to the horizon of the swirling extreme binary system. This is a sort of rotational Meissner effect\footnote{The Meissner effect for extremal black hole immersed in Bonnor-Melvin external magnetic field was shown in \cite{magnetised-kerr-cft}.} which was already observed for accelerating swirling black holes in \cite{marcoa-remove}.\\
Then, if the swirling effect is not relevant in this extremal near-horizon regime, we may hope to detect some similarity with the standard Kerr black hole. Indeed, inspecting the equations (\ref{near-1}) and \ref{near-2}) we observe that these expressions resemble the standard near horizon of a single extremal Kerr geometry \cite{strominger-kerr-cft}, whose values are
\bea
        \G_0(y) &=& \hat{a}^2 (1+y^2) \hspace{2 cm}\s_0(y)  = 1 \nn \\
        \psi_0(y)  &=&  \frac{2 \sqrt{1-y^2}}{1+y^2}    \hspace{2.3cm} \k_0 = 1  \ .
\eea
The similitude between the two metrics is so stringent that we can map the near horizon geometry of the first regularisation\footnote{Strictly speaking the bottom black hole of the first regularisation and the top of the second. The other two black holes have the same near-horizon geometry but they are rotating in opposite directions, so they have $\k$ switched.} just by
\beq \label{map1}
      ^{(1)} \hat{a} \ \longmapsto \  \frac{4 \sqrt{2} a^4 M^2}{(a^2+M^2)^{3/2}|M^2 + 2aM - a^2|} \ ;
\eeq
while the near horizon geometry of the binary system with the second regularisation matches the Kerr one when  
\beq \label{map2}
      ^{(2)} \hat{a} \ \longmapsto \  \frac{4 \sqrt{2} a^4 M^2}{(a^2+M^2)^{3/2}(a^2 + 2aM - M^2)} \ .
\eeq
Since these black holes are extremal we can use these mappings to check the masses and angular momenta computed in section \ref{sec:thermo}. In fact, the extremal Kerr black hole mass and angular momentum are 
\beq
           m_0 = \hat{a}   \ , \hspace{3cm} J_0 = \hat{a}^2
\eeq
Thanks to the maps in (\ref{map1}), (\ref{map2}) we can infer the mass and angular momentum of each constituent of the binary system for the first and the second distinct regularisations, i.e
\beq
          ^{(1)}\bar{m}_\pm  =  \frac{4 \sqrt{2} a^4 M^2}{(a^2+M^2)^{3/2}|M^2 + 2aM - a^2|}  \ , \hspace{1.5cm} ^{(1)}\bar{J}_\pm =  \frac{\mp 32 a^8 M^4}{(a^2+M^2)^3(M^2 + 2aM - a^2)^2} 
\eeq
and
\beq
          ^{(2)}\bar{m}_\pm  =  \frac{4 \sqrt{2} a^4 M^2}{(a^2+M^2)^{3/2}(a^2 + 2aM - M^2)}   \ , \hspace{1.5cm} ^{(2)}\bar{J}_\pm =  \frac{\pm 32 a^8 M^4}{(a^2+M^2)^3(a^2 + 2aM - M^2)^2}
\eeq
The difference of positivity on the angular momentum basically comes from the signs of $^{(i)}\k$ in (\ref{near-1}), (\ref{near-2}) and reflects the counterrotating nature of the binary system.  \\
Note that these values, $^{(i)}m_\pm , \ ^{(i)}J_\pm$, for the mass and angular momentum  of each black hole correspond exactly to the ones computed in the previous section, as can be seen in (\ref{m-tilde}) and (\ref{m-J}).  So this is a good sanity check, not only for the extremal configuration, but it represents a good limiting test for the non-extremal masses and angular momenta. 

Clearly when the binary metric is not regularised, that is when $\D_\varphi$ and $\jmath$ are not set as in (\ref{j-Dphi}), the analogy with the single asymptotically flat and rotating black hole is not possible because the near horizon geometry of a singular black hole cannot be mapped into the one of a regular one. Thus, the similitude between the mass and angular momentum of the two different systems does not hold either. Nevertheless, it is possible to build a map between the near horizon metric of the two extremal irregular black holes and the near horizon geometry of the accelerating Kerr metric. To do so, we just set the value of the gauge constant $\D_\varphi=\bar{\D}_\varphi$ to erase one of the two conical singularities of each event horizon, while leaving $\jmath$ free, for details see appendix \ref{app:irregular_NH}. A similar behaviour was previously observed also in \cite{marcoa-remove}. We conjecture that, at least in general relativity, independently on the nature of the conical singularities, at extremality all near horizon geometries of single black holes endowed with angular defects can be cast into the near horizon metric of the accelerating Kerr solution, or Kerr-Newman whether the Maxwell electromagnetic field is included.  \\

\section{Conclusions}

In this article we presented a new analytical and exact solution in pure four-dimensional general relativity which describes a couple of counterrotating extremal Kerr black holes embedded into a rotating background, known as swirling universe.\\
We have shown that the spin-spin interaction between the angular momenta of the black holes can balance the gravitational attraction, so the unphysical strings or the rods which usually affect the seed metric can be removed. This means that the counter-rotating binary black hole admits an equilibrium configuration without external fields or energy momentum contributions to the Einstein field equations. This result clarifies a long standing open problem in general relativity. \\
Our analysis pointed out that it is fundamental that the black holes carry angular momentum. Indeed the immersion of the double Schwarzschild solution into the rotating background does not switch on the spin-spin interaction, even though the metric is rotating. That's because the components of the Bach-Weyl metric into the swirling universe have no intrinsic angular momentum, even though it is rotating.

Remarkably we have shown an explicit example where the standard Einstein's theory for gravitation admits repulsive interaction between ordinary matter, thus without violations of energy conditions and keeping the spacetime manifold completely void from conical or curvature\footnote{The Ehlers transformation in four-dimensional general relativity (\ref{ehlers-transf}) is known not to introduce curvature singularities when applied to a black hole metric. Anyway the Kretschmann scalar invariant has been analysed for a wide range of parameters and its plot, which confirms it is bounded, is portrayed in appendix \ref{app:kret}.} singularities outside the event horizons, even asymptomatically. 

These outcomes do not stem from any kind of approximation but from considering the full general relativity contribution. Actually our finding contradicts the common gravitational spin-spin knowledge based on approximations \cite{wald}, because we found that the repulsion effect, which is able to balance the system, is present in the case of antiparallel angular momenta too. It would be interesting to check, outside the approximation approach, whether also similar equilibrium configurations of black holes are allowed when the spins are parallel instead of anti-parallel. This case should be favoured according to the estimate (\ref{spin-spin-wald}).

No hair theorems or uniqueness theorems for non-singular black hole configurations are circumvented by the presence of the rotating background that modifies the asymptotic behaviour of the metric field, which is not Minkowskian any more.

The duality between the constituents of the binary black hole system and a two-dimensional boundary conformal field theory is verified near the event horizons. Actually a parametric map between each black binary component and the Kerr metric confirms also a kind of rotational Meissner effect. In fact the extremal near horizon geometry coincides with the near horizon metric approximation of extreme standard Kerr, therefore the external rotational contribution is negligible, in this regime.   

Even though we presented a model only for extreme twin black holes, just for the sake of simplicity, all we discussed in this article can be straightforwardly generalised for the non-extremal case or for black holes with different masses or angular momenta. Actually a more general setting would improve the phenomenological properties of the system. Furthermore an extra unconstrained parameter, such as $\jmath$, would allow one to discuss a possible topological phase transition of the event horizon, when increasing the background angular velocity, between an overstretched oblate single black hole versus a binary black hole layout\footnote{A preliminary study of this mechanism has been proposed in \cite{mati}, however a larger parametric space would significantly improve the picture.}.

It would be also interesting to study the possible stability of these black binary equilibrium configurations, for both feasible regularisations. 

In principle, the same technique for embedding black holes into the rotating universe has been established also for charged black holes \cite{marcoa-remove}; therefore also Kerr-Newman black holes can be included in this picture. Clearly, a composition endowed with repulsive electric charges can improve the balance conditions and refine the present model.\\

\paragraph{Acknowledgements}
{\small  This article is based on the results of the bachelor thesis \cite{mati} by M.T. A Mathematica notebook containing the main solution presented in this article can be found in the arXiv source folder. %This work has been partially funded by INFN and by MIUR-PRIN contract 2017CC72MK-003} %n$^\textrm{o}$
}\\

\appendix

\section{The Bach-Weyl black hole binary in the swirling universe}
\label{app:bach-weyl-swirling}

Embedding a couple of coaxial static black holes into the swirling universe \cite{swirling} makes the two constituents stationary (counter) rotating. However the two sources do not acquire the angular momenta needed to trigger the spin-spin interaction. Therefore this solution cannot enjoy equilibrium configuration between the two black holes. Here we show, in detail, why this model is too simple to be regularised from axial conical singularities. \\
For the construction of the metric we start by considering, as a seed, the static binary found by Bach and Weyl \cite{bach-weyl} 
\beq
\label{bach-weyl}
          ds^2 = - \frac{\mu_1\mu_3}{\mu_2\mu_4} \ dt^2 + \frac{16\ C_f \ \mu_1^3 \mu_2^5 \mu_3^3 \mu_4^5 \ (d\r^2+dz^2)}{\mu_{12}^2 \mu_{14}^2 \mu_{23}^2 \mu_{34}^2 W_{13}^2 W_{24}^2 W_{11} W_{22} W_{33} W_{44}} + \rho^2 \frac{\mu_2 \mu_4}{\mu_1 \mu_3} \ d\varphi^2 \ ,
\eeq
where
\begin{equation}
\begin{split}
\label{eqn:BW_solitons}
    \mu_i(\r,z) = w_i - z + \sqrt{\rho^2 + (z - w_i)^2} \; , \qquad \mu_{ij} = (\mu_i - \mu_j)^2 \; ,\qquad W_{ij} = \rho^2 + \mu_i\,\mu_j \; ,
\end{split}
\end{equation}
are the basic building blocks of the solitonic inverse scattering technique. 

\begin{figure}[h]
%	\captionsetup[subfigure]{labelformat=empty}
	\centering
%	\hspace{-0.2cm}
	\includegraphics[scale=0.38]{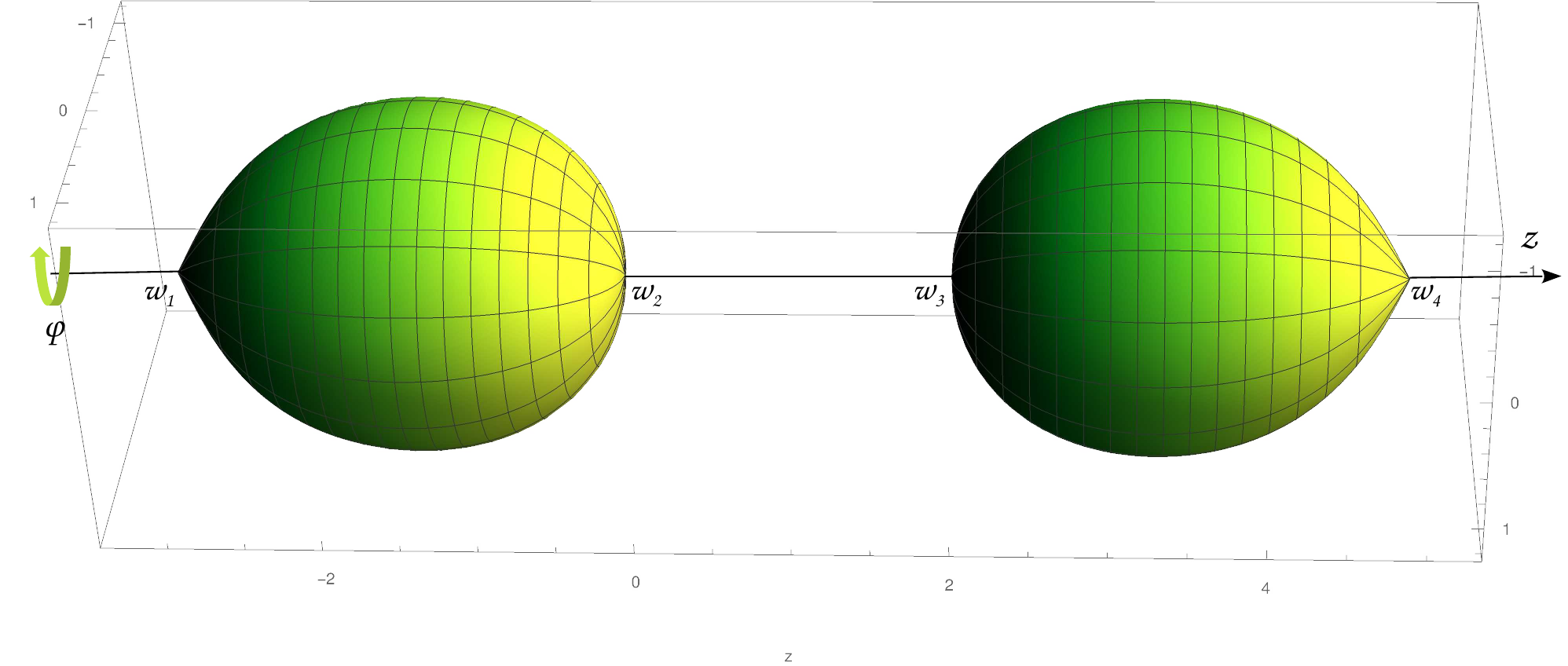}
	\caption{\small Isometric embedding of the event horizons of a Bach Weyl metric immersed into the swirling universe into the three dimensional Euclidean flat space, for $\jmath=1/5 , \ w_1=-2, \ w_2=-1, \ w_3=1, \ w_4 =2 $. Here the gauge freedom in $C_f$ is fixed to remove the conical singularity between the two black holes. Nevertheless conical singularities,  interpreted as semi-infinite strings, are present on the axis of symmetry, in the external region to balance an equilibrium configuration.}
	\label{fig:BW}
\end{figure}

This metric describes a couple of Schwarzschild black holes, located at ($\r=0, \ w_1<z<w_2$) and ($\r=0, \ w_3<z<w_4$), kept apart by a rod of repulsive matter between them or two semi-infinite axial strings, which extend to infinity. In the first case the delta-like energy momentum tensor modelling the rod violates all the physical energy conditions, while in the second case the energy momentum tensor becomes infinite, because the string has strictly positive energy density and unbounded length. To cure these issues, we try to add a rotating background, such as the swirling universe, thanks to the Ehlers transformation (\ref{ehlers-transf}) of the Ernst equations. First of all we need to extract, just by comparison, the seed Ernst potential for the seed metric (\ref{bach-weyl}) in terms of the Lewis-Weyl-Papapetrou metric (\ref{lwpm}), that is
\beq
          \Er =   -\rho^2 \, \frac{\mu_2 \, \mu_4}{\mu_1 \, \mu_3} \; .
\eeq
Then thanks to the Ehlers transformation (\ref{ehlers-transf}) we can generate the new solution, which in terms of the Ernst potentials reads
\beq
          \hat{\Er} = \frac{-\m_2\m_4 \r^2}{\m_1\m_3- i \jmath \m_2 \m_4 \r^2} \ .
\eeq
The metric representation of the above GR vacuum solution is reached using the Ernst potential definition (\ref{Ernst-pot}) to get 
\beq
       \hat{f}(\r,z) = - \frac{\r^2\m_1 \m_2 \m_3 \m_4}{\m_1^2 \m_3^2 + \jmath^2 \r^4 \m_2^2 \m_4} 
\eeq
and integrating the eq. (\ref{h}) to obtain
\beq
     \hat{\omega}(\r,z) = 2 \jmath (\m_2-w_2+\m_4-w_4-\m_3+w_3-\m_1+w_1) + \om_0
\eeq
The metric, in the LWP form (\ref{lwpm}), is completely determined, since the gamma function remains the same as the seed
\beq
        \hat{\gamma}(\r,z) = \frac{16 C_f \r^2 \m_1^2 \m_2^6 \m_3^2 \m_4^6}{\m_{12} \m_{23} \m_{14} \m_{34} W_{11} W_{22} W_{33} W_{44} W_{13}^2 W_{24}^2 } \ .
\eeq
Clearly, this stationary rotating metric describes two Schwarzschild black holes embedded into the swirling universe; however this spacetime possesses no angular momentum (as the single Schwarzschild in the swirling universe \cite{swirling}). This is the reason why the spin-spin interaction is not present and we cannot regularise all the conical singularities on the axis of symmetry.\\
In fact, computing the conicities, as in section \ref{sec:equilibrium}, in the three sectors outside the black holes' horizons we get \\
 \hspace{1.2cm} $\bullet$ for \(z < w_1\) or \(z > w_4\): 
\beq \label{BW-conext}
            \lim_{\rho \to 0}\frac{L}{R} =  \ \lim_{\r \to 0} \frac{2\pi}{\r} \sqrt{\frac{g_{\varphi\varphi}}{g_{zz}}} \ =\ 2\pi \frac{4 (w_1 - w_2) (w_2 - w_3) (w_1 - w_4) (w_3 - w_4)}{\sqrt{C_f}} \ ,
\eeq
$\bullet$ for \(w_1 < w_2 < z < w_3 < w_4\):
\beq\label{BW-conint}
 \lim_{\rho \to 0}\frac{L}{R} =  \ \lim_{\r \to 0} \frac{2\pi}{\r} \sqrt{\frac{g_{\varphi\varphi}}{g_{zz}}} \ = \ 2 \pi \frac{4 (w_1 - w_2) (w_1 - w_3) (w_2 - w_4) (w_3 - w_4)}{\sqrt{C_f}} \ .
\eeq
The above two quantities do not depend on the swirling parameter $\jmath$, so they coincide with the seed conicities,  therefore there are no hopes to regularise the system without spoiling the black hole interpretation. In fact, imposing the regularity conditions that both (\ref{BW-conext}) $=$ (\ref{BW-conint}) = $2\pi$, implies imposing the two constraint on the parameters, for instance $C_f$ and $w_4$. While the first parameter is basically free (apart from its sign to ensure the proper metric signature), the constraint on the second implies that $ w_4 < w_3 $, which is against the model hypothesis. Hence only the constraint on $C_f$ can be enforced to regularise either the external regions or the internal one. 
From picture \ref{fig:BW} it is possible to appreciate the latter option: the two black holes are kept apart by two semi-infinite strings on the axis of symmetry, while the region between the horizons is free of conical singularity because $C_f$ is fixed as follows
\beq
           C_f = 16 (w_1 - w_2)^2 (w_1 - w_3)^2 (w_2 - w_4)^2 (w_3 - w_4)^2 \ .
\eeq
  
Further informations about physical properties, conserved charges and thermodynamics for this system can be found in \cite{mati}.\\

\subsection{Limit to the swirling Schwarzschild black hole}

Considering that the two black holes of masses $m_1$, $m_2$ are centred in $z_1$, $z_2$ respectively, and that the event horizons are defined by 
\beq
       w_1 = z_1-m_1 \ , \hspace{1.5cm} w_2 = z_1 + m_1  , \hspace{1.5cm} w_3 = z_2 - m_2  , \hspace{1.5cm} w_4 = z_2 + m_2
\eeq 
we can rescale $C_f$ and take the limit for $z_1 \to -\infty$ to remove one of the two constituents to get 
\beq
       ds^2 = \frac{\m_3 \m_4 \r^2}{\m_3^2+\jmath^2 \r^4 \m_4^2} \Big[d\varphi - 2\jmath(2z+\m_4-\m_3+\tilde{\om}_0) dt  \Big]^2 - \frac{\m_3^2+\jmath^2 \r^4 \m_4^2}{\m_3 \m_4 \r^2} \left[ \r^2 dt^2 + \frac{16 \tilde{C}_f \r^2 \m_4^4}{\m_{34} W_{33} W_{44}} (d\r^2+dz^2)  \right] \ . \label{Sch-Swi-rho-z}
\eeq

This solution represents the Schwarzschild black hole embedded in the swirling universe. Even though the black hole is not centred in the origin of the coordinates $z_2 \neq 0$, but it can be adjusted in any point of the axis of symmetry, we can show that this is exactly the same solution found in \cite{swirling}. Actually, thanks to the change of coordinates
\beq
             \r = \sqrt{r^2-2mr} \sin \theta \ , \hspace*{1.3cm} z = z_2 + (r-m) \cos \theta \ , \hspace*{1.3cm} m_2 = m \ , \hspace{1.3cm} \tilde{C}_f = -m^2 \ , \nn
\eeq
the metric (\ref{Sch-Swi-rho-z}) can be cast into spherical coordinates:
\beq \label{sch-swirl}
      ds^2 =(1+\jmath^2 r^4 \sin^4 \theta)\left[ - \left(1-\frac{2m}{r}\right) dt^2 + \frac{dr^2}{1-\frac{2m}{r}} +r^2 d\theta^2 \right] + \frac{r^2 \sin^2 \theta}{1+\jmath^2 r^4 \sin^4 \theta} \Big[ d\varphi - [4\jmath(r-2m)\cos \theta + \tilde{\tilde{\om}}_0 ] dt \Big]^2  ,
\eeq
with $\tilde{\tilde{\om}}_0 = 2 \jmath(2m+2z_2+\tilde{\om}_0)$. This coincides, up to a redefinition of the gauge constant $\tilde{\tilde{\om}}_0$, with the metric studied in \cite{swirling}. So moving the black hole along the z-axis by changing $z_2$ corresponds only in the shifting, by a constant factor, the angular speed $\tilde{\tilde{\om}}_0$ of an asymptotically far observer. So basically the position of the black hole on the axis of symmetry can be adjusted by a gauge transformation, i.e. properly tuning $\tilde{\tilde{\om}}_0$. \\
Clearly, when $\jmath = 0$ the metric (\ref{sch-swirl}) becomes the usual Schwarzschild black hole, while for $m=0$ it recovers the swirling background.\\

\section{$\hat{\om}(x,y)$}
\label{app:omega}

\beq
            \hat{\om}(x,y) = \om(x,y) + \frac{\jmath \, \hat{\om}_1(x,y) + \jmath^2 \, \hat{\om}_2(x,y)}{\hat{D}(x,y)}
\eeq

\beq
\begin{split}
\hspace{-10pt}
    \hat{\om}_1(x,y) &= \,4 \,M \, \left(a^2+M^2\right) \, y \,  \Big[6 \left(x^2-1\right)^3 \left(x^2+3\right) \left(a^{12}-M^{12}\right)+x \left(x^2-1\right)^3 \left(x^2+15\right) \left(a^{12}+M^{12}\right) \\
    &+ 4 \, a^6 \,  M^6 x \, [5 x^8-12 x^6 \left(2 y^2+1\right)+2 x^4 \left(6 y^4+40 y^2-39\right)+x^2 \left(8 \left(-4 y^4+y^2+5\right) y^2+428\right)\\
    & \qquad \qquad \quad +8 y^8-36 y^4-15]  \\
    &+2 a^4 M^4 \left(a^4-M^4\right) \big[15 x^8+x^6 \left(8-72 y^2\right)+2 x^4 \left(12 y^4+68 y^2-189\right)+ 24 x^2 \left(-2 y^6+4 y^4+5 y^2+6\right) \\
    &\qquad \qquad \qquad \qquad \quad+8 y^2 \left(-2 y^4+y^2+1\right)-45\big] \\
    &+a^4 M^4 \left(a^4+M^4\right) x\, \left[15 x^8-4 x^6 \left(16 y^2+3\right)+6 x^4 \left(16 y^4+7\right)-4 x^2 \left(16 y^6+267\right)+16 y^8+15\right] \\
    &+2 a^2 M^2 \big[\left(a^8+M^8\right) x\, \left(4 \left(9 x^4-2 x^2+9\right) y^4-8 x^2 \left(x^4+10 x^2+5\right) y^2+3 \left(x^2+1\right) \left(x^6+3 x^4+23 x^2+5\right)\right) \\
    & \qquad \ + \left(a^8-M^8\right) 4\, \left(3 x^8+x^6+54 x^4-27 x^2+\left(15 x^4-6 x^2-1\right) y^4-\left(x^2+1\right) \left(9 x^4+14 x^2+1\right) y^2+9\right)\big]\Big] \nn
\end{split}
\eeq

\beq
\begin{split}
\hspace{-13pt}
    \hat{\om}_2(x,y) &=\, 8 \, \frac{M^2  \left(a^2+M^2\right)^2}{a} \, y \, \Big[4 \left(3 x^8-30 x^4-5\right) \left(a^{12}+M^{12}\right)+2 x \left(x^8+12 x^6-42 x^4-20 x^2-15\right) \left(a^{12}-M^{12}\right) \\
    &+ 2 a^6 M^6 [-69 x^8+444 x^6-1026 x^4+84 x^2+4 \left(x^4-6 x^2-3\right) y^8+8 \left(x^6+3 x^4+15 x^2-3\right) y^6 \\
    &\qquad \qquad +\left(-9 x^8+44 x^6-198 x^4+204 x^2+215\right) y^4+2 \left(15 x^8-8 x^6-186 x^4+176 x^2+3\right) y^2-41]  \\
    &+ 2 a^4 M^4 \big[\left(a^4-M^4\right) \, x \, (5 x^8-156 x^6+630 x^4+32 \left(x^2+3\right) y^6-332 x^2-8 \left(3 x^6-9 x^4+29 x^2+9\right) y^4 \\
    &\qquad \qquad+ 16 \left(3 x^6+3 x^4-25 x^2-5\right) y^2+45)  \\
    &\qquad \qquad- \left(a^4+M^4\right) 2 \,(x^8 \left(3 y^4-10 y^2+18\right)-2 x^6 \left(y^6-5 y^4+15 y^2+5\right) \\
    &\qquad \qquad- x^4 \left(y^8+6 y^6-42 y^4-10 y^2+219\right)+ 6 x^2 y^2 \left(y^6-5 y^4-y^2+21\right)+3 y^8+6 y^6+63 y^4-5)\big] \\
    &+ 8 a^2 M^2 \left(a^8-M^8\right) \big[x^9-3 x^7 \left(y^4-2 y^2+5\right)-3 x^5 \left(y^4-10 y^2+15\right)+x^3 \left(-25 y^4+50 y^2+43\right) \\
    &\qquad \qquad \qquad \qquad \quad+5 x y^2 \left(3 y^2+2\right)\big] \\
    &+ a^2 M^2 \left(a^8+M^8\right) \big[9 x^8-404 x^6+246 x^4-84 x^2-3 \left(x^8+28 x^6-10 x^4+60 x^2-15\right) y^4 \\
    &\qquad \qquad \qquad \qquad \quad +2 \left(5 x^8+68 x^6+166 x^4+84 x^2-3\right) y^2+41\big]\Big]  \nn
\end{split}
\eeq

\beq
\begin{split}
\hspace{-25pt}
    \hat{D}(x,y) &= \left(M^2-a^2\right) \Big[a^{12} (x+1)^8+M^{12} (x-1)^8 \\
    &+ 4 a^6 M^6 \big[5 x^8-4 x^6 \left(6 y^2+7\right)+6 x^4 \left(2 y^4+24 y^2+11\right)-4 x^2 \left(8 y^6+6 y^4+78 y^2+17\right)+8 y^8 \\
    &\qquad \qquad -100 y^4+96 y^2+1\big] \\
    &+8 a^4 M^4 x \left(a^4-M^4\right) \big[5 x^6-21 x^4+8 \left(x^2+3\right) y^4+27 x^2-24 \left(x^4-4 x^2+1\right) y^2-16 y^6+5\big] \\
    &+ a^4 M^4 \left(a^4+M^4\right) \big[15 x^8-4 x^6 \left(16 y^2+7\right)+x^4 \left(96 y^4-70\right)+x^2 \left(-64 y^6+768 y^2+228\right) \\
    &\qquad \qquad \qquad \qquad+ 16 y^2 \left(y^6+16 y^2-16\right)-1\big] \\
    &+ 2 a^2 M^2 \big[\left(a^8-M^8\right) \left(16 x^7-48 x^5 y^2+80 x^3 y^4-192 x^3 y^2-96 x^3-48 x y^4+48 x y^2-16 x\right) \\
    &\qquad \qquad+\left(a^8+M^8\right)(3 x^8-8 x^6 y^2+28 x^6+36 x^4 y^4-144 x^4 y^2-66 x^4+24 x^2 y^4-72 x^2 y^2 \\
    &\qquad \qquad \qquad \qquad \qquad -60 x^2-28 y^4+32 y^2-1)\big]\Big] \nn
\end{split}
\eeq
\\

\section{Majumdar-Papapetrou black holes have infinite proper distance}
\label{app:MP}

The Majumdar-Papapetrou solution \cite{majumdar}, \cite{papapetrou}, \cite{hartle-hawking} generically describes a collection of extremal black holes at equilibrium. The gravitational force is perfectly balanced by the Maxwell repulsive electromagnetic interaction. Therefore there are no conical (and no curvature) singularities outside the event horizons. Here we consider the minimal spacetime configuration composed by just an axisymmetric binary: the two sources are located on the z-axis, at $z_1$ and $z_2$. It is described by the following metric
\beq
        ds^2 = - \left(1 + \frac{q_1}{x_1(\r,z)} + \frac{q_2}{x_2(\r,z)} \right)^{-2} dt^2 + \left(1 + \frac{q_1}{x_1(\r,z)} + \frac{q_2}{x_2(\r,z)} \right)^{2} \left( d\r^2 +dz^2 + \r^2 d\varphi^2 \right) \ , \nn
\eeq
where
\beq
           x_i(\r,z) := \sqrt{\r^2+(z-z_i)^2} \ ,\nn
\eeq
supported by the electric potential 
\beq
            A_t = \left(1 + \frac{q_1}{x_1(\r,z)} + \frac{q_2}{x_2(\r,z)} \right)^{-1} \ .      \nn
\eeq
While the coordinate distance $ |z_2-z_1| $, between the constituent of the charged binary is not divergent, the proper distance between the two black holes' horizons goes to infinity:
\beq
        d_p \ = \ \lim_{\r \to 0} \int_{z_1}^{z_2} \sqrt{g_{zz}(\r,z)} \ dz  \ = \ \int_{z_1}^{z_2} \left( 1 + \frac{q_1}{z-z_1} + \frac{q_2}{z_2-z}\right) dz \ \to \ \infty \ \  .
\eeq
This is a typical feature of extremal horizons, not necessarily binary systems. Because of the degeneracy between the inner and the outer horizons, the proper distance between the extreme event horizon and another spacetime point is unbounded. This fact also explains the reason why extremal black holes do not radiate and have zero temperature.   \\

\section{Near horizon geometry with a conical singularity}
\label{app:irregular_NH}

Here we show the similitude between the irregular near horizon geometry of the binary system and a prototypical conical extreme rotating black hole horizon, such as the accelerating extreme Kerr. Without losing any generality we suppose that the gauge freedom in the range of the azimuthal coordinate (encoded in the parameter $\tilde{\D}_\varphi$) has been fixed to remain with only one conical defect, on the south pole.\\  
The near horizon metric of the extreme accelerating Kerr solution (NHEAK) can be described analogously, to (\ref{twisted}), by

\beq \label{ds-near}
          ds^2 = \tilde{\G}(\tilde{y}) \left[ - r^2 d\tau^2 +\frac{dr^2}{r^2} + \tilde{\s}(\tilde{y})^2 \frac{d\tilde{y}^2}{1-\tilde{y}^2} + \tilde{\psi}(\tilde{y}) \Big( d\varphi^2 - \tilde{\k} r d\tau \Big)^2 \right] \ ,
\eeq
%\MA{Io qua, per massima chiarezza e anche a costo di sembrare pedante a qualcuno, inserirei proprio i valori che descrivono l'andamento near horizon come nella referenza, cioe' le formule (3.4)-(3.6) di https://arxiv.org/pdf/1605.06131 o (5.1)-(5.3) di https://arxiv.org/pdf/2207.14305 credo}
with, as in \cite{c-cft},\footnote{Function \(\tilde{\k}\) is presented with an opposite sign with respect to what is shown in \cite{c-cft}: this is justified by the presence of an overall minus sign before \(\tilde{\k}\) itself in metric (\ref{twisted}), which was instead reabsorbed in the function in \cite{c-cft}.} 
\bea 
    \tilde{\G}(\tilde{y}) &=&  \frac{\tilde{a}^2\,(1 + \tilde{y}^2)}{(1-\tilde{a}^2\tilde{A}^2)\,(1 + \tilde{a}\tilde{A} \, \tilde{y})^2} \ ,
    \hspace{3.3 cm}  \tilde{r}_0 \ = \ \sqrt{\frac{2 \, \tilde{a}^2}{1 - \tilde{a}^2 \tilde{A}^2}} \ ,\nn \\ 
   \tilde{\psi}(\tilde{y})  &=&  \frac{2\, \tilde{\D}_\varphi \, \tilde{a}^2 \, \sqrt{1- \tilde{y}^2}}{\tilde{\G}(\tilde{y}) \, \sqrt{1-\tilde{a}^2\tilde{A}^2} \, (1 + \tilde{a}\tilde{A} \, \tilde{y})} \ ,\hspace{2.8cm} \tilde{\k} \ = \ \frac{\tilde{r}_0^2}{2 \, \tilde{a}^2\,\tilde{\D}_\varphi} \ , \label{functions-near} \\
    \tilde{\s}(\tilde{y})  &=&   \pm \frac{\sqrt{1-\tilde{a}^2\tilde{A}^2}}{1 + \tilde{a}\tilde{A} \, \tilde{y}}\ ,  \hspace{4.6cm}  \tilde{\D}_\varphi \ = \ \frac{1}{1 + 2 \, \tilde{a}\tilde{A} + \tilde{a}^2 \tilde{A}^2} \ .\nn 
\eea
\\
In order to fully comprehend the similitude between the geometries, we can perform a coordinate transformation
%\MT{may be redundant and quite confusing at this point, so it may be better not to mention this change; or write it better} \MA{si, qua non si puo usare x, che e' gia impegnata nel paper, puoi usare forse $\tilde{y}$ o $ \mathring{y}$, pensado le (5.1)-(5.3)di 2207.14305 scritte in termini della stessa coordinata}
\beq
    \tilde{y}(y) = - \frac{\tilde{a}\tilde{A} + y}{1 + \tilde{a}\tilde{A} y} \ ,
\eeq
that allows to reabsorb the value of \(\Tilde{\s}(y)\) as to reduce it to the unity. This process leads to the following expressions: 
%\bea
%            \G_\pm(y) &=& \frac{M^2}{(a^2+M^2)^3}\,\big[A\,(y^2 + 1) \pm B\, y\big] \hspace{1.8 cm} \s(y)  = 1 \nn \\ 
%            \psi_\pm(y)  &=&  \frac{16 \, a^4 \, M^4  \, \bar{\D}_\varphi \, (1-y^2)}{\G^2_\pm \; (a^2+M^2)^2}    \hspace{3.4cm} \k_\pm = \, \pm \frac{2 \,M \, C}{a \, (a^2 + M^2)\,\bar{\D}_\varphi}  \ ,
%\eea

%where
%\begin{equation}
%\begin{split}
%    A(a,M) = \; &a^2 \left(a^4+6 a^2 M^2+M^4\right)+8\, a \, M \left(a^6-3 a^4 M^2-5 a^2 M^4-M^6\right) \, j \, + \\
%    &+16\, M^2 \left(a^8+4 a^6 M^2+6 a^4 M^4+4 a^2 M^6+M^8\right) \, j^2 \\
%    B(a,M) = \; &2 a^2 \left(a^4-2 a^2 M^2+M^4\right)+16\,a\, M \left(a^6+5 a^4 M^2+3 a^2 M^4-M^6\right)\,j + \\
%    &+32 \,M^2 \left(a^8-4 a^6 M^2-10 a^4 M^4-4 a^2 M^6+M^8\right)\,j^2 \\
%    C(a,M) = \; &a^2 -8 a M^3 \,j +16 \, M^2 (M^4 - a^4)\,j^2  \; .
%\end{split}
%\end{equation}

\bea\label{nheak}
           \tilde{\G}(y) &=&  \,\frac{\tilde{a}^2 [1 + 4\tilde{a}\tilde{A} y + y^2 + \tilde{a}^2\tilde{A}^2 (1 + y^2)]}{(1-\tilde{a}^2 \tilde{A}^2)^3} \ , \nn \\ 
            \tilde{\psi}(y)  &=& \frac{2  \, \tilde{a}^2 \, \tilde{\D}_\varphi \, \sqrt{1-y^2}}{(1-\tilde{a}^2 \tilde{A}^2) \; \tilde{\G}(y)} \ ,\\
            \tilde{\s}(y)  &=& 1  \; . \nn
\eea

We choose \(\bar{\D}_\varphi\), from eq. (\ref{cony}), to remove only the singularities in \(z < -k\) and \(z > k\), while keeping an angular defect in the region between the black holes (i.e. $-k<z<k$) for the bottom black hole, that is
\beq
 \bar{\D}_\varphi = \frac{\left(4 a^2 j M+a-4 j M^3\right)^2}{a^2} \ .
\eeq
 On the other hand we have to regularise the internal region $-k<z<k$ for the event horizon locate at positive $z$ by considering
\beq
      \bar{\D}_\varphi =\frac{4 M^2 \left(-4 a^2 j M+a-4 j M^3\right)^2}{\left(a^2+M^2\right)^2} \ ,
\eeq
in order to remain with a smooth north pole on both constituents. Hence, by comparison, we can map the near horizon metric of the irregular binary constituents (\ref{Gamma-psi-inizio}), (\ref{Gamma-psi-fine}) into the NHEAK (\ref{nheak}) by setting

\bea
    \tilde{a}_\pm &=& \frac{16\, \sqrt{\mp \,a^3 M^5 \, (-a^2+8 a M^3 \, j +16 a^4 M^2 j^2-16  M^6 \, j^2)^3}}{\left[4 a^4 j M+a^3 \left(1 \mp 8 j M^2\right) \pm 2 a^2 M+a \left(M^2 \mp 8 j M^4\right)-4 j M^5\right]^2} \; ,\nn \\
    \\
    \tilde{A}_\pm &=& \frac{1}{\tilde{a}_\pm} \left[ \frac{4 a M \left(4 a^2 j M-a+4 j M^3\right)}{4 a^4 j M + a^3 \left(1 \mp 8 j M^2 \right) \pm 2 a^2 M+a \left(M^2 \mp 8 j M^4\right)-4 j M^5}\pm 1 \right] .\nn
\eea

Thus, at extemality, the near horizon geometry of black holes with a conical singularity seems to display universal properties, in fact they are diffeomorphic. As shown in this example, and also in \cite{marcoa-remove}, it is not relevant if the angular defect is caused by an acceleration due to a Rindler horizon, a gravitational spin-spin interaction or by the presence of another gravitational source, all these cases can be cast into (\ref{ds-near})-(\ref{functions-near}).  \\

\newpage

\section{Kretschmann scalar invariant}
\label{app:kret}

\begin{figure}[h]
%	\captionsetup[subfigure]{labelformat=empty}
	\centering
%	\hspace{-0.2cm}
	\includegraphics[scale=0.6]{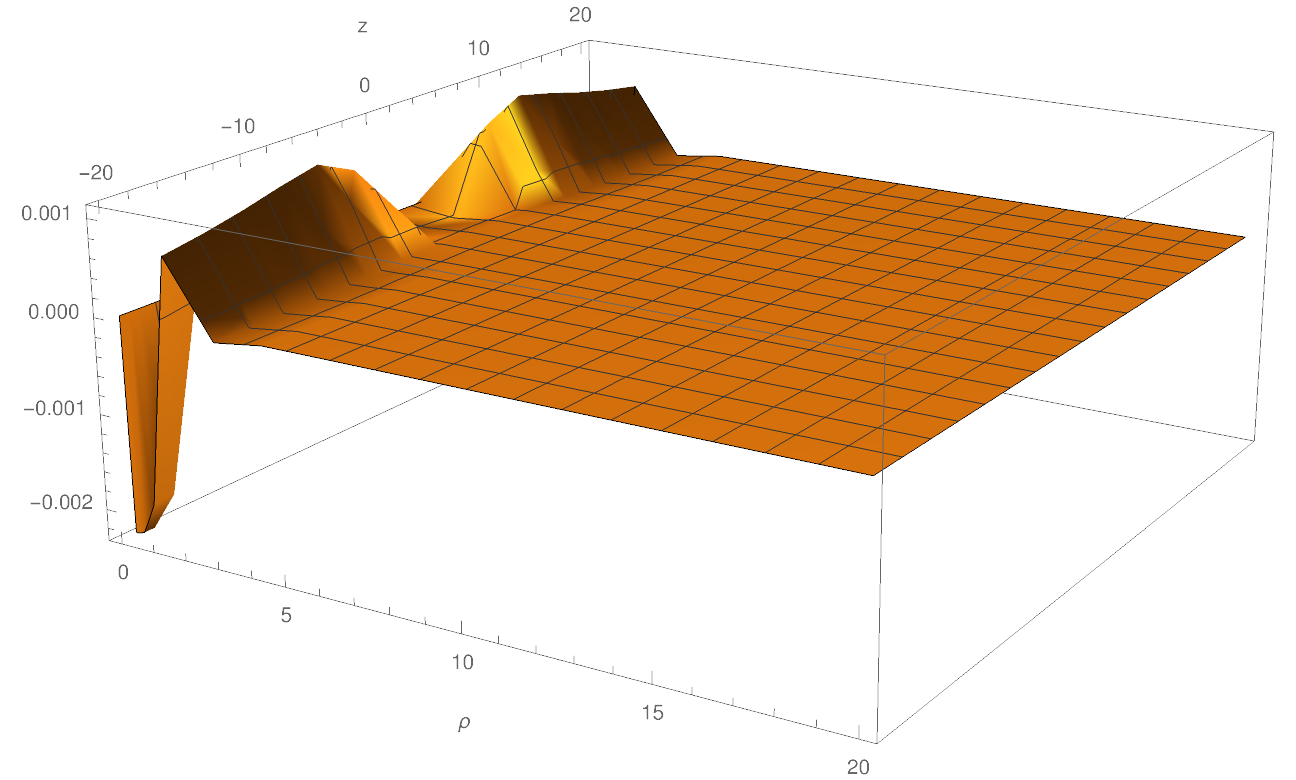}
	\caption{\small Kretschmann scalar invariant for $M=1$, $a=2$ and ($\jmath$, $\bar{\Delta}_\varphi$) as in (\ref{j-Dphi}) is limited.}
	\label{fig:kret}
\end{figure}

To confirm that the metric is not affected by curvature singularities in the domain of outer communication, we present an example for the finiteness of the Kretschmann scalar invariant $R_{\m\n\s\l} R^{\m\n\s\l}$. We use the ($\r,z$) coordinates which do not cover the spacetime inside the event horizons. As can be seen from figure \ref{fig:kret}, the curvature invariant does not diverge anywhere. 

\newpage







\begin{thebibliography}{99}


\bibitem{KramerNeug}
 D.~Kramer, G.~Neugebauer,
  {\it ``The superposition of two Kerr solutions''},
   \href{https://www.sciencedirect.com/science/article/abs/pii/0375960180905563}{Physics Letters A \textbf{75} (1980), no.4, 259--261}.

\bibitem{dietz}
 W. Dietz and  C. Hoenselaers, {\it ``Two mass solutions of Einstein's vacuum equations: The double Kerr solution"}, \href{https://doi.org/10.1016/0003-4916(85)90301-X}{Annals of Physics {\bf 165} , no. 2 (1985): 319-383}.
 
\bibitem{wald}
R.~M.~Wald,
{\it ``Gravitational spin interaction''},
 \href{https://doi.org/10.1103/PhysRevD.6.406}{Phys. Rev. D \textbf{6} (1972), 406-413}.
%doi:10.1103/PhysRevD.6.406
%211 citations counted in INSPIRE as of 16 Nov 2024


\bibitem{Herdeiro:2008kq}
C.~A.~R.~Herdeiro and C.~Rebelo,
{\it ``On the interaction between two Kerr black holes''},
 \href{https://doi.org/10.1088/1126-6708/2008/10/017}{JHEP \textbf{10} (2008), 017};
\href{https://arxiv.org/pdf/0808.3941}{\tt[arXiv:0808.3941 [gr-qc]]}.
%33 citations counted in INSPIRE as of 21 Sep 2024

\bibitem{swirling}
 M.~Astorino, R.~Martelli and A.~Vigan\`o,
  {\it ``Black holes in a swirling universe''},
   \href{https://doi.org/10.1103/PhysRevD.106.064014}{Phys. Rev. D \textbf{106} (2022) no.6, 064014};
    \href{https://arxiv.org/pdf/2205.13548}{\tt [arXiv:2205.13548 [gr-qc]]}

\bibitem{harrison}
B. Kent Harrison, {\it ``New Solutions of the Einstein‐Maxwell Equations from Old''}, 
\href{https://doi.org/10.1063/1.1664508}{J. Math. Phys. 9 (11): 1744–1752, (1968)}


\bibitem{Moreira:2024sjq}
Z.~S.~Moreira, C.~A.~R.~Herdeiro and L.~C.~B.~Crispino,
{\it ``Twisting shadows: Light rings, lensing, and shadows of black holes in swirling universes''}, 
\href{https://doi.org/10.1103/PhysRevD.109.104020}{Phys. Rev. D \textbf{109} (2024) no.10, 104020}; 
\href{https://arxiv.org/pdf/2401.05658}{\tt [arXiv:2401.05658 [gr-qc]]}.
%10 citations counted in INSPIRE as of 09 Mar 2025


\bibitem{ernst-magnetic} 
  F.~J.~Ernst,
   {\it ``Black holes in a magnetic universe''},
    \href{https://doi.org/10.1063/1.522781}{J.\ Math.\ Phys.\  {\bf 17}, no. 1, 54 (1976).}

%\bibitem{marcoa-equivalence}
%M.~Astorino,
%{\it ``Equivalence principle and generalised accelerating black holes from binary systems''},
%\href{https://doi.org/10.1103/PhysRevD.109.084038}{Phys. Rev. D \textbf{109} (2024) no.8, 8};
%\href{https://arxiv.org/pdf/2312.00865.pdf}{\tt [arXiv:2312.00865 [gr-qc]]}.

\bibitem{majumdar}
S.~D.~Majumdar,
{\it ``A class of exact solutions of Einstein's field equations''},
\href{https://doi.org/10.1103/PhysRev.72.390}{Phys. Rev. \textbf{72} (1947), 390-398}
%doi:10.1103/PhysRev.72.390
%457 citations counted in INSPIRE as of 20 Jul 2023

 
\bibitem{papapetrou} 
  A. Papapetrou {\it ``A static solution of the equations of the gravitational field for an arbitary charge-distribution''}, Proceedings of the Royal Irish Academy. Section A: Mathematical and Physical Sciences. Vol. 51. Royal Irish Academy, (1945) 191; \href{https://www.jstor.org/stable/20488481}{\tt https://www.jstor.org/stable/20488481}

\bibitem{hartle-hawking}
  J.~B.~Hartle and S.~W.~Hawking,
  {\it ``Solutions of the Einstein-Maxwell equations with many black holes''},
  Commun.\ Math.\ Phys.\  {\bf 26} (1972) 87.\ \ 
  \href{https://link.springer.com/article/10.1007%2FBF01645696}{\tt doi:10.1007/BF01645696}
  %%CITATION = doi:10.1007/BF01645696;%%
  %222 citations counted in INSPIRE as of 30 May 2019

\bibitem{marcoa-binary}
 M.~Astorino and A.~Vigan\`o,
  {\it ``Binary black hole system at equilibrium''},
    \href{https://doi.org/10.1016/j.physletb.2021.136506}{Phys. Lett. B \textbf{820} (2021), 136506};
     %doi:10.1016/j.physletb.2021.136506
      \href{https://arxiv.org/pdf/2104.07686}{\tt [2104.07686 [gr-qc]]}.
       %4 citations counted in INSPIRE as of 13 Mar 2022

\bibitem{many-rotating}
 M.~Astorino and A.~Vigan\`o,
  {\it ``Charged and rotating multi-black holes in an external gravitational field''},
   \href{https://doi.org/10.1140/epjc/s10052-021-09693-6}{Eur. Phys. J. C \textbf{81} (2021) no.10, 891}; 
    \href{https://arxiv.org/abs/2105.02894}{\tt [arXiv:2105.02894 [gr-qc]]}.
     %10 citations counted in INSPIRE as of 17 Jun 2023


\bibitem{multipolar-acc}
 M.~Astorino and A.~Vigan\`o,
  {\it``Many accelerating distorted black holes''},
   \href{https:/Debever:1983pi/doi.org/10.1140/epjc/s10052-021-09693-6}{Eur. Phys. J. C \textbf{81} (2021) no.10, 891};
    %doi:10.1140/epjc/s10052-021-09693-6
     \href{https://arxiv.org/pdf/2106.02058.pdf}{\tt [2106.02058 [gr-qc]]}.
      %6 citations counted in INSPIRE as of 05 Apr 2022

\bibitem{bubble}
 M.~Astorino, R.~Emparan and A.~Vigan\`o,
 {\it ``Bubbles of nothing in binary black holes and black rings, and viceversa''},
  \href{https://doi.org/10.1007/JHEP07(2022)007}{JHEP \textbf{07} (2022), 007} ;
   \href{https://arxiv.org/pdf/2204.09690.pdf}{\tt [arXiv:2204.09690 [hep-th]]}.


\bibitem{Kastor-MP}
D.~Kastor and J.~H.~Traschen,
{\it ``Cosmological multi - black hole solutions''},
\href{https://doi.org/10.1103/PhysRevD.47.5370}{Phys. Rev. D \textbf{47} (1993), 5370-5375};
%doi:10.1103/PhysRevD.47.5370
\href{https://arxiv.org/pdf/hep-th/9212035}{\tt [arXiv:hep-th/9212035 [hep-th]]}.
%239 citations counted in INSPIRE as of 16 Nov 2024

\bibitem{Klemm-MP}
S.~Chimento and D.~Klemm,
{\it ``Multicentered black holes with a negative cosmological constant''},
\href{https://doi.org/10.1103/PhysRevD.89.024037}{Phys. Rev. D \textbf{89} (2014) no.2, 024037};
%doi:10.1103/PhysRevD.89.024037
\href{https://arxiv.org/pdf/1311.6937}{\tt [arXiv:1311.6937 [hep-th]]}.
%12 citations counted in INSPIRE as of 16 Nov 2024


\bibitem{emparan-dihole}
R.~Emparan,
{\it ``Black diholes''}, 
\href{https://doi.org/10.1103/PhysRevD.61.104009}{Phys. Rev. D \textbf{61} (2000), 104009}
\href{https://arxiv.org/pdf/hep-th/9906160}{\tt [arXiv:hep-th/9906160 [hep-th]]}.
%73 citations counted in INSPIRE as of 02 Nov 2024
    

\bibitem{emparan-teo}
R.~Emparan and E.~Teo,
{\it ``Macroscopic and microscopic description of black diholes''},
\href{https://doi.org/10.1016/S0550-3213(01)00319-4}{Nucl. Phys. B \textbf{610} (2001), 190-214} ; 
\href{https://arxiv.org/pdf/hep-th/0104206}{\tt [arXiv:hep-th/0104206 [hep-th]]}.
%52 citations counted in INSPIRE as of 11 Jan 2021

\bibitem{Emparan:2001gm}
R.~Emparan and M.~Gutperle,
{\it ``From p-branes to fluxbranes and back''},
\href{https://doi.org/10.1088/1126-6708/2001/12/023}{JHEP \textbf{12} (2001), 023};
%doi:10.1088/1126-6708/2001/12/023
\href{https://arxiv.org/pdf/hep-th/0111177}{\tt [arXiv:hep-th/0111177 [hep-th]]}.
%53 citations counted in INSPIRE as of 29 Jan 2025

\bibitem{Dias:2023rde}
 O.~J.~C.~Dias, G.~W.~Gibbons, J.~E.~Santos and B.~Way,
 {\it ``Static Black Binaries in de Sitter Space''}, 
\href{https://doi.org/10.1103/PhysRevLett.131.131401}{Phys. Rev. Lett. \textbf{131} (2023) no.13, 131401};
    \href{https://arxiv.org/pdf/2303.07361}{\tt [arXiv:2303.07361 [gr-qc]]}

\bibitem{Dias:2024dxg}
 O.~J.~C.~Dias, J.~E.~Santos and B.~Way,
  {\it ``Spinning Black Binaries in de Sitter space''},
   \href{https://doi.org/10.1103/PhysRevLett.133.191401}{Phys. Rev. Lett. \textbf{133} (2024) no.19, 191401}; 
    \href{https://arxiv.org/pdf/2406.10333}{\tt [arXiv:2406.10333 [gr-qc]]}
  
\bibitem{Herdeiro:2023mpt}
 C.~A.~R.~Herdeiro and E.~Radu,
  {\it ``Two Schwarzschild-like black holes balanced by their scalar hair''},
   \href{https://doi.org/10.1103/PhysRevD.107.064044}{Phys. Rev. D \textbf{107} (2023) no.6, 064044};
    %doi:10.1103/PhysRevD.107.064044
     \href{https://arxiv.org/pdf/2302.00016}{\tt [arXiv:2302.00016 [gr-qc]]}.
      %6 citations counted in INSPIRE as of 21 Sep 2024


\bibitem{Herdeiro:2023roz}
 C.~A.~R.~Herdeiro and E.~Radu,
  {\it ``Two Spinning Black Holes Balanced by Their Synchronized Scalar Hair''},
\href{https://doi.org/10.1103/PhysRevLett.131.121401}{Phys. Rev. Lett. \textbf{131} (2023) no.12, 121401};
  \href{https://arxiv.org/pdf/2305.15467}{\tt [arXiv:2305.15467 [gr-qc]]}
     %11 citations counted in INSPIRE as of 21 Sep 2024

\bibitem{ernst1}
   F.~J.~Ernst,
  {\it ``New formulation of the axially symmetric gravitational field problem''},
  \href{https://doi.org/10.1103/PhysRev.167.1175}{Phys.\ Rev.\  {\bf 167} (1968) 1175}.

\bibitem{enhanced}
  M.~Astorino,
  {\it ``Enhanced Ehlers Transformation and the Majumdar-Papapetrou-NUT Spacetime''},
  \href{https://doi.org/10.1007/JHEP01(2020)123}{JHEP \textbf{01} (2020), 123}; 
   %doi:10.1007/JHEP01(2020)123
   \href{https://arxiv.org/pdf/1906.08228}{\tt [arXiv:1906.08228 [gr-qc]]}


\bibitem{PD-NUTs}
  M.~Astorino and G. Boldi,
     {\it ``Plebanski-Demianski goes NUTs (to remove the Misner string)''},
      \href{https://doi.org/10.1007/JHEP08(2023)085}{JHEP \textbf{08} (2023), 085};
       \href{https://arxiv.org/pdf/2305.03744.pdf}{\tt [arXiv:2305.03744 [gr-qc]]}


\bibitem{bach-weyl}
 R. Bach and H. Weyl,
  {\it ``Neue l{\"o}sungen der einsteinschen gravitationsgleichungen''},
   \href{https://doi.org/10.1007/BF01485284}{Mathematische Zeitschrift 13, 134–145 (1922)}.


\bibitem{manko}
V.~S.~Manko, E.~D.~Rodchenko, E.~Ruiz and B.~I.~Sadovnikov,
{\it ``Exact solutions for a system of two counter-rotating black holes''},
\href{https://doi.org/10.1103/PhysRevD.78.124014}{Phys. Rev. D \textbf{78} (2008), 124014}
\href{https://arxiv.org/pdf/0809.2422}{\tt [arXiv:0809.2422 [gr-qc]]}
%20 citations counted in INSPIRE as of 21 Sep 2024


\bibitem{marcoa-remove}
 M.~Astorino,
  {\it ``Removal of conical singularities from rotating C-metrics and dual CFT entropy''}
   \href{https://doi.org/10.1007/JHEP10(2022)074}{JHEP \textbf{10} (2022), 074};
    \href{https://arxiv.org/pdf/2207.14305}{\tt [arXiv:2207.14305 [gr-qc]]}

\bibitem{tomimatsu-84}
A.~Tomimatsu,
{\it ``Distorted Rotating black holes''}, \href{https://lib-extopc.kek.jp/preprints/PDF/1984/8404/8404024.pdf}{\tt RRK 84-7};
\href{https://doi.org/10.1016/0375-9601(84)90134-8}{Phys. Lett. A \textbf{103} - 8 (1984), 374-376}

\bibitem{kodama}
H.~Kodama and W.~Hikida,
{\it ``Global structure of the Zipoy-Voorhees-Weyl spacetime and the delta=2 Tomimatsu-Sato spacetime''},
\href{https://doi.org/10.1088/0264-9381/20/23/011}{Class. Quant. Grav. \textbf{20} (2003), 5121-5140};
 \href{https://arxiv.org/pdf/gr-qc/0304064}{\tt [arXiv:gr-qc/0304064 [gr-qc]]}.
%56 citations counted in INSPIRE as of 05 Nov 2024

\bibitem{smarr-embedding}
L.~Smarr,
{\it ``Surface Geometry of Charged Rotating Black Holes''},
\href{https://doi.org/10.1103/PhysRevD.7.289}{Phys. Rev. D \textbf{7} (1973), 289-295}.
%doi:10.1103/PhysRevD.7.289
%123 citations counted in INSPIRE as of 29 Jan 2025

\bibitem{strominger-kerr-cft}
M.~Guica, T.~Hartman, W.~Song and A.~Strominger,
{\it ``The Kerr/CFT Correspondence''},
\href{https://doi.org/10.1103/PhysRevD.80.124008}{Phys. Rev. D \textbf{80} (2009), 124008} ;
%doi:10.1103/PhysRevD.80.124008
\href{https://arxiv.org/pdf/0809.4266}{\tt [arXiv:0809.4266 [hep-th]]}.%[arXiv:0809.4266 [hep-th]].
%722 citations counted in INSPIRE as of 23 Jun 2022

\bibitem{strominger-duals}
T.~Hartman, K.~Murata, T.~Nishioka and A.~Strominger,
{\it ``CFT Duals for Extreme Black Holes''},
\href{https://doi.org/10.1088/1126-6708/2009/04/019}{JHEP \textbf{04} (2009), 019} ;
%doi:10.1088/1126-6708/2009/04/019
\href{https://arxiv.org/pdf/0811.4393}{\tt [arXiv:0811.4393 [hep-th]]}.%[arXiv:0811.4393 [hep-th]].
%219 citations counted in INSPIRE as of 23 Jun 2022

\bibitem{c-cft}
M.~Astorino,
{\it ``CFT Duals for Accelerating Black Holes''}
\href{https://doi.org/10.1016/j.physletb.2016.07.019}{Phys. Lett. B \textbf{760} (2016), 393-405} ;
%doi:10.1016/j.physletb.2016.07.019
\href{https://arxiv.org/pdf/1605.06131}{\tt [arXiv:1605.06131 [hep-th]]}.
%32 citations counted in INSPIRE as of 23 Jun 2022

\bibitem{magnetised-RN-cft}
M.~Astorino,
{\it ``Microscopic Entropy of the Magnetised Extremal Reissner-Nordstrom Black Hole''},
\href{https://doi.org/10.1007/JHEP10(2015)016}{JHEP \textbf{10} (2015), 016} ;
\href{https://arxiv.org/pdf/1507.04347}{\tt [arXiv:1507.04347 [hep-th]]}.
%18 citations counted in INSPIRE as of 23 Jun 2022

\bibitem{magnetised-kerr-cft}
M.~Astorino,
{ \it ``Magnetised Kerr/CFT correspondence''},
\href{https://doi.org/10.1016/j.physletb.2015.10.017}{Phys. Lett. B \textbf{751} (2015), 96-106};
\href{https://arxiv.org/pdf/1508.01583}{\tt [arXiv:1508.01583 [hep-th]]}.
%31 citations counted in INSPIRE as of 23 Jun 2022


\bibitem{compere-kerr-cft}
G.~Comp\`ere,
{\it ``The Kerr/CFT correspondence and its extensions''},
\href{https://doi.org/10.1007/s41114-017-0003-2}{Living Rev. Rel. \textbf{15} (2012), 11};
%doi:10.1007/s41114-017-0003-2
\href{https://arxiv.org/pdf/1203.3561}{\tt [arXiv:1203.3561 [hep-th]]}.%[arXiv:1203.3561 [hep-th]].
%197 citations counted in INSPIRE as of 23 Jun 2022

\bibitem{lucietti-kunduri}
H.~K.~Kunduri, J.~Lucietti and H.~S.~Reall,
{\it ``Near-horizon symmetries of extremal black holes''},
\href{https://doi.org/10.1088/0264-9381/24/16/012}{Class. Quant. Grav. \textbf{24} (2007), 4169-4190} ;
%doi:10.1088/0264-9381/24/16/012
\href{https://arxiv.org/pdf/0705.4214}{\tt [arXiv:0705.4214 [hep-th]]}.%[arXiv:0705.4214 [hep-th]].
%282 citations counted in INSPIRE as of 23 Jun 2022

\bibitem{mati}
M.~Torresan,
 {\it ``Gravitational spin-spin interaction and binary black hole system at equilibrium''},
 Università degli Studi di Milano (2024); 
 \href{https://doi.org/10.6084/m9.figshare.28514441.v1}{\tt https://doi.org/10.6084/m9.figshare.28514441.v1}






%%%%%%%%%%%%%%%%%%%%%%%%%%%%%%%%%%%%%%%%%%%%%%%%%%%%%%%%%%%%%%%%%%%%%%%%%%%%%%%%%%%%%%%%%%%%%%%%%%%%%%%%%%%%%%%%%%%%%%




\iffalse


\bibitem{Debever}
R. Debever, 
{\it On type D expanding solutions of Einstein–Maxwell equations}, 
\href{https://doi.org/10.1016/0375-9601(83)90469-3}{Bull. Soc. Math.
Belg. 23 (1971) 360–76}.

\bibitem{Plebanski-Demianski}
  J.~F.~Plebanski and M.~Demianski,
  {\it ``Rotating, charged, and uniformly accelerating mass in general relativity''},
   \href{https://doi.org/10.1016/0003-4916(76)90240-2}{Annals Phys. \textbf{98} (1976), 98-127}
    %509 citations counted in INSPIRE as of 18 Jan 2023


\bibitem{mann-stelea-chng}
B.~Chng, R.~B.~Mann and C.~Stelea,
{\it ``Accelerating Taub-NUT and Eguchi-Hanson solitons in four dimensions''},
  \href{https://doi.org/10.1103/PhysRevD.74.084031}{Phys. Rev. D \textbf{74} (2006), 084031};
   \href{https://arxiv.org/pdf/gr-qc/0608092.pdf}{\tt [arXiv:gr-qc/0608092]}
    %12 citations counted in INSPIRE as of 18 Jan 2023


\bibitem{tesi-giova}
  G.~Boldi,
   {\it ``Ehlers transformation and accelerating spacetimes with a gravomagnetic monopole''},
    \href{https://inspirehep.net/literature/2652409}{Università degli Studi di Milano (2022)}


\bibitem{PD-NUTs}
  M.~Astorino and G. Boldi,
     {\it ``Plebanski-Demianski goes NUTs (to remove the Misner string)''},
      \href{https://doi.org/10.1007/JHEP08(2023)085}{JHEP \textbf{08} (2023), 085};
       \href{https://arxiv.org/pdf/2305.03744.pdf}{\tt [arXiv:2305.03744 [gr-qc]]}


\bibitem{Type-I}
M.~Astorino,
 {\it ``Accelerating and Charged Type I Black Holes''},
  \href{https://doi.org/10.1103/PhysRevD.108.124025}{Phys. Rev. D \textbf{108} (2023) no.12, 124025};
   \href{https://arxiv.org/pdf/2307.10534.pdf}{\tt [arXiv:2307.10534 [gr-qc]]}


\bibitem{Podolsky-nut}
J.~Podolsky and A.~Vratny,
{\it ``Accelerating NUT black holes''}, 
  \href{https://doi.org/10.1103/PhysRevD.102.084024}{Phys. Rev. D \textbf{102} (2020) no.8, 084024}; 
   \href{https://arxiv.org/pdf/2007.09169.pdf}{\tt [arXiv:2007.09169 [gr-qc]]}
    %9 citations counted in INSPIRE as of 18 Jan 2023

\bibitem{New-Look}
J.~B.~Griffiths and J.~Podolsky,
{\it ``A New look at the Plebanski-Demianski family of solutions''},
\href{https://doi.org/10.1142/S0218271806007742}{Int. J. Mod. Phys. D \textbf{15} (2006), 335-370};
\href{https://arxiv.org/pdf/gr-qc/0511091.pdf}{\tt [arXiv:gr-qc/0511091 [gr-qc]]}
%198 citations counted in INSPIRE as of 02 Apr 2024

\bibitem{Podolsky-New-Lambda}
J.~Podolsky and A.~Vratny,
{\it ``New form of all black holes of type D with a cosmological constant''},
 \href{https://doi.org/10.1103/PhysRevD.107.084034}{Phys. Rev. D \textbf{107} (2023) no.8, 084034}; 
\href{https://doi.org/10.1103/PhysRevD.108.129902}{[erratum: Phys. Rev. D \textbf{108} (2023) no.12, 129902]};
\href{https://arxiv.org/pdf/2212.08865.pdf}{\tt[arXiv:2212.08865 [gr-qc]]}
%8 citations counted in INSPIRE as of 16 Mar 2024

\bibitem{Debever:1983pi}
R.~Debever, N.~Kamran and R.~g.~Mclenaghan,
{\it ``A single expression for the general solution of Einstein's vacuum and electrovac field equations with cosmological constant for Petrov type D admitting a non-singular aligned Maxwell field''},
\href{https://doi.org/10.1016/0375-9601(83)90469-3}{Phys. Lett. A \textbf{93} (1983), 399-402}
%6 citations counted in INSPIRE as of 04 Aug 2024

\bibitem{marcoa-equivalence}
M.~Astorino,
{\it ``Equivalence principle and generalised accelerating black holes from binary systems''},
\href{https://arxiv.org/pdf/2312.00865.pdf}{\tt [arXiv:2312.00865 [gr-qc]]}



\bibitem{enhanced}
  M.~Astorino,
  {\it ``Enhanced Ehlers Transformation and the Majumdar-Papapetrou-NUT Spacetime''},
  \href{https://doi.org/10.1007/JHEP01(2020)123}{JHEP \textbf{01} (2020), 123}; 
   %doi:10.1007/JHEP01(2020)123
   \href{https://arxiv.org/pdf/1906.08228}{\tt [arXiv:1906.08228 [gr-qc]]}


\bibitem{charging}
M.~Astorino,
{\it ``Charging axisymmetric space-times with cosmological constant''},
\href{https://doi.org/10.1007/JHEP06(2012)086}{JHEP \textbf{06} (2012), 086} ;
\href{https://arxiv.org/pdf/1205.6998.pdf}{\tt [arXiv:1205.6998 [gr-qc]]}
%35 citations counted in INSPIRE as of 27 Sep 2022


\bibitem{stephani-big-book}
  H.~Stephani, D.~Kramer, M.~A.~H.~MacCallum, C.~Hoenselaers and E.~Herlt,
  {``Exact solutions of Einstein's field equations''}, \ 
  \href{https://doi.org/10.1017/CBO9780511535185}{\tt [doi:10.1017/CBO9780511535185]}

\bibitem{OPA}
H. Ovcharenko, J. Podolsky and M. Astorino,
{\it ``Black holes of type D revisited:
relating their various metric forms''}, work in progress




\bibitem{ernst-magnetic} 
  F.~J.~Ernst,
   {\it ``Black holes in a magnetic universe''},
    \href{https://doi.org/10.1063/1.522781}{J.\ Math.\ Phys.\  {\bf 17}, no. 1, 54 (1976).}


\bibitem{ernst-remove}
  F.~J.~Ernst,
   {\it ``Removal of the nodal singularity of the C-metric''}, 
    \href{http://scitation.aip.org/content/aip/journal/jmp/17/4/10.1063/1.522935}{J. Math. Phys. {\bf 17}, 515 (1976)}.


\bibitem{swirling}
 M.~Astorino, R.~Martelli and A.~Vigan\`o,
  {\it ``Black holes in a swirling universe''},
   \href{https://doi.org/10.1103/PhysRevD.106.064014}{Phys. Rev. D \textbf{106} (2022) no.6, 064014};
    \href{https://arxiv.org/pdf/2205.13548}{\tt [arXiv:2205.13548 [gr-qc]]}.
     %0 citations counted in INSPIRE as of 26 Jun 2022


\bibitem{marcoa-remove}
 M.~Astorino,
  {\it ``Removal of conical singularities from rotating C-metrics and dual CFT entropy''}
   \href{https://doi.org/10.1007/JHEP10(2022)074}{JHEP \textbf{10} (2022), 074};
    \href{https://arxiv.org/pdf/2207.14305}{\tt [arXiv:2207.14305 [gr-qc]]}.


\bibitem{emparan-reall}
 R.~Emparan and H.~S.~Reall,
  {``Generalized Weyl solutions''}, 
   \href{https://doi.org/10.1103/PhysRevD.65.084025}{Phys. Rev. D \textbf{65} (2002), 084025}; 
    \href{https://arxiv.org/pdf/hep-th/0110258.pdf}{\tt [arXiv:hep-th/0110258 [hep-th]]}.
     %280 citations counted in INSPIRE as of 07 Oct 2023


\bibitem{bach-weyl}
 R. Bach and H. Weyl,
  {\it ``Neue l{\"o}sungen der einsteinschen gravitationsgleichungen''},
   \href{https://doi.org/10.1007/BF01485284}{Mathematische Zeitschrift 13, 134–145 (1922)}.

\bibitem{Wang:1996sn}
 Y.~c.~Wang,
  {\it ``Vacuum C metric and the metric of two superposed Schwarzschild black holes''},
   \href{https://doi.org/10.1103/PhysRevD.55.7977}{Phys. Rev. D \textbf{55} (1997), 7977-7979}
    %doi:10.1103/PhysRevD.55.7977
     %7 citations counted in INSPIRE as of 25 Oct 2023

\bibitem{Klemm-embedding}
 A.~Gnecchi, K.~Hristov, D.~Klemm, C.~Toldo and O.~Vaughan,
  {\it ``Rotating black holes in 4d gauged supergravity''}
   \href{https://doi.org/10.1007/JHEP01(2014)127}{JHEP \textbf{01} (2014), 127};
    \href{https://arxiv.org/pdf/1311.1795.pdf}{\tt [arXiv:1311.1795 [hep-th]]}.
     %97 citations counted in INSPIRE as of 24 Oct 2023


\bibitem{ernst-generalized-c} 
 F.~J.~Ernst,
  {\it ``Generalized C-metric''},
   \href{https://doi.org/10.1063/1.523896}{J.\ Math.\ Phys.\  {\bf 19}, 1986-1987 (1978).}

   
\bibitem{many-rotating}
 M.~Astorino and A.~Vigan\`o,
  {\it ``Charged and rotating multi-black holes in an external gravitational field''},
   \href{https://doi.org/10.1140/epjc/s10052-021-09693-6}{Eur. Phys. J. C \textbf{81} (2021) no.10, 891}; 
    \href{https://arxiv.org/abs/2105.02894}{\tt [arXiv:2105.02894 [gr-qc]]}.
     %10 citations counted in INSPIRE as of 17 Jun 2023


\bibitem{marcoa-binary}
 M.~Astorino and A.~Vigan\`o,
  {\it ``Binary black hole system at equilibrium''},
    \href{https://doi.org/10.1016/j.physletb.2021.136506}{Phys. Lett. B \textbf{820} (2021), 136506};
     %doi:10.1016/j.physletb.2021.136506
      \href{https://arxiv.org/pdf/2104.07686}{\tt [2104.07686 [gr-qc]]}.
       %4 citations counted in INSPIRE as of 13 Mar 2022


\bibitem{deCastro}
 G.~M.~de Castro and P.~S.~Letelier,
  {\it ``Black holes surrounded by thin rings and the stability of circular orbits''},
   \href{https://doi.org/10.1088/0264-9381/28/22/225020}{Class. Quant. Grav. \textbf{28} (2011), 225020}


\bibitem{ernst2}
 F.~J.~Ernst,
  {\it ``New Formulation of the Axially Symmetric Gravitational Field Problem. II''},
    \href{https://doi.org/10.1103/PhysRev.168.1415}{\tt Phys.\ Rev.\  {\bf 168} (1968) 1415.}


\bibitem{enhanced}
  M.~Astorino,
  {\it ``Enhanced Ehlers Transformation and the Majumdar-Papapetrou-NUT Spacetime''},
  \href{https://doi.org/10.1007/JHEP01(2020)123}{JHEP \textbf{01} (2020), 123}; 
   %doi:10.1007/JHEP01(2020)123
   \href{https://arxiv.org/pdf/1906.08228}{\tt [arXiv:1906.08228 [gr-qc]]}.
    

\bibitem{belinski-book}
 V. Belinski, E. Verdaguer, \href{https://doi.org/10.1017/CBO9780511535253}{\it ``Gravitational solitons''}, Cambridge, Cambridge Univ. Press, 2001.



\bibitem{Podolsky-2021}
 J.~Podolsky and A.~Vratny,
  {\it ``New improved form of black holes of type D''},
    \href{https://doi.org/10.1103/PhysRevD.104.084078}{Phys. Rev. D \textbf{104} (2021), 084078};
     %doi:10.1103/PhysRevD.104.084078
      \href{https://arxiv.org/pdf/2108.02239}{\tt [arXiv:2108.02239 [gr-qc]]}.
       %9 citations counted in INSPIRE as of 27 Mar 2023

\bibitem{hong-teo-rotating-C}
 K.~Hong and E.~Teo,
  {\it ``A New form of the rotating C-metric''},
B. Kent Harrison; New Solutions of the Einstein‐Maxwell Equations from Old. J. Math. Phys. 1 November 1968; 9 (11): 1744–1752. https://doi.org/10.1063/1.1664508
   \href{https://doi.org/10.1088/0264-9381/22/1/007}{Class. Quant. Grav. \textbf{22} (2005), 109-118};
    \href{https://arxiv.org/pdf/gr-qc/0410002.pdf}{\tt [arXiv:gr-qc/0410002 [gr-qc]]}.
     %55 citations counted in INSPIRE as of 07 Jan 2024

\bibitem{melvin-lambda}
M.~Astorino,
{\it ``Charging axisymmetric space-times with cosmological constant''},
\href{https://doi.org/10.1007/JHEP06(2012)086}{JHEP \textbf{06} (2012), 086} ;
\href{https://arxiv.org/pdf/1205.6998.pdf}{\tt [arXiv:1205.6998 [gr-qc]]}.
%35 citations counted in INSPIRE as of 27 Sep 2022


%%%%%%%%%%%%%%%%%%%%%%%%%%%%%%%%%%%%%%%%%%%%%%%%%%%%%%%%%%%%%%%%%%%%%%%%%%%%%%%%%%%%%%%%%%%%%%%%%%%%%%%%%%%%%%%%%%%%%%%%








\bibitem{Podolsky-nut}
J.~Podolsky and A.~Vratny,
{\it ``Accelerating NUT black holes''}, 
 \href{https://doi.org/10.1103/PhysRevD.102.084024}{Phys. Rev. D \textbf{102} (2020) no.8, 084024}; 
\href{https://arxiv.org/pdf/2007.09169.pdf}{\tt [arXiv:2007.09169 [gr-qc]]}. 
%9 citations counted in INSPIRE as of 18 Jan 2023


\bibitem{Plebanski-Demianski}
  J.~F.~Plebanski and M.~Demianski,
  {\it ``Rotating, charged, and uniformly accelerating mass in general relativity''},
  \href{https://doi.org/10.1016/0003-4916(76)90240-2}{Annals Phys. \textbf{98} (1976), 98-127}.
  %509 citations counted in INSPIRE as of 18 Jan 2023

\bibitem{tesi-giova}
  G.~Boldi,
    {\it ``Ehlers transformation and accelerating spacetimes with a gravomagnetic monopole''},
     \href{https://inspirehep.net/literature/2652409}{Università degli Studi di Milano (2022)}

\bibitem{PD-NUTs}
  M.~Astorino and G. Boldi,
     {\it ``Plebanski-Demianski goes NUTs (to remove the Misner string)''},
     \href{https://doi.org/10.1007/JHEP08(2023)085}{JHEP \textbf{08} (2023), 085};
 \href{https://arxiv.org/pdf/2305.03744.pdf}{\tt [arXiv:2305.03744 [gr-qc]]}.
  
  
\bibitem{ernst-remove}
  F.~J.~Ernst,
  {\it ``Removal of the nodal singularity of the C-metric''}, 
   \href{http://scitation.aip.org/content/aip/journal/jmp/17/4/10.1063/1.522935}{J. Math. Phys. {\bf 17}, 515 (1976)}.


\bibitem{ernst-generalized-c} 
  F.~J.~Ernst,
  {\it ``Generalized C-metric''},
   \href{https://doi.org/10.1063/1.523896}{J.\ Math.\ Phys.\  {\bf 19}, 1986-1987 (1978).}


%\cite{Astorino:2021rdg}
\bibitem{multipolar-acc}
M.~Astorino and A.~Vigan\`o,
{\it``Many accelerating distorted black holes''},
\href{https:/Debever:1983pi/doi.org/10.1140/epjc/s10052-021-09693-6}{Eur. Phys. J. C \textbf{81} (2021) no.10, 891};
%doi:10.1140/epjc/s10052-021-09693-6
\href{https://arxiv.org/pdf/2106.02058.pdf}{\tt [2106.02058 [gr-qc]]}.
%6 citations counted in INSPIRE as of 05 Apr 2022

\bibitem{ernst2}
  F.~J.~Ernst,
  {\it ``New Formulation of the Axially Symmetric Gravitational Field Problem. II''},
  \href{https://doi.org/10.1103/PhysRev.168.1415}{\tt Phys.\ Rev.\  {\bf 168} (1968) 1415.}

\bibitem{ernst-hauser}
I.~Hauser and F.~J.~Ernst,
{\it ``Proof of a generalized Geroch conjecture for the hyperbolic Ernst equation''},
\href{https://doi.org/10.1023/A:1002701301339}{Gen. Rel. Grav. \textbf{33} (2001), 195-293}; 
%doi:10.1023/A:1002701301339
\href{https://arxiv.org/pdf/gr-qc/0002049.pdf}{\tt [arXiv:gr-qc/0002049 [gr-qc]]}.
%15 citations counted in INSPIRE as of 11 Jul 2023

\bibitem{many-rotating}
M.~Astorino and A.~Vigan\`o,
{\it ``Charged and rotating multi-black holes in an external gravitational field''},
\href{https://doi.org/10.1140/epjc/s10052-021-09693-6}{Eur. Phys. J. C \textbf{81} (2021) no.10, 891}, 
\href{https://arxiv.org/abs/2105.02894}{\tt [arXiv:2105.02894 [gr-qc]]}.
%10 citations counted in INSPIRE as of 17 Jun 2023

\bibitem{reina-treves}
  A.~Reina and A. Treves
  {\it ``NUT-like generalization of axisymmetric gravitational fields''} ,  \href{https://doi.org/10.1063/1.522614}{Journal of Mathematical Physics 16, 834 (1975)}.

\bibitem{enhanced}
  M.~Astorino,
  {\it ``Enhanced Ehlers Transformation and the Majumdar-Papapetrou-NUT Spacetime''},
  \href{https://doi.org/10.1007/JHEP01(2020)123}{JHEP \textbf{01} (2020), 123} ; 
   %doi:10.1007/JHEP01(2020)123
   \href{https://arxiv.org/pdf/1906.08228}{\tt [arXiv:1906.08228 [gr-qc]]}.
    %3 citations counted in INSPIRE as of 17 Jan 2022
    
\bibitem{ernst-magnetic} 
  F.~J.~Ernst,
  {\it ``Black holes in a magnetic universe''},
  \href{https://doi.org/10.1063/1.522781}{J.\ Math.\ Phys.\  {\bf 17}, no. 1, 54 (1976).}

\bibitem{marcoa-pair}
  M.~Astorino,
  {\it ``Pair Creation of Rotating Black Holes''},
  \href{https://doi.org/10.1103/PhysRevD.89.044022}{Phys. Rev. D \textbf{89} (2014) no.4, 044022};
  %doi:10.1103/PhysRevD.89.044022
  \href{https://arxiv.org/pdf/1312.1723.pdf}{\tt [arXiv:1312.1723~[gr-qc]]}.
  %9 citations counted in INSPIRE as of 20 May 2022

\bibitem{swirling}
M.~Astorino, R.~Martelli and A.~Vigan\`o,
{\it ``Black holes in a swirling universe''},
\href{https://doi.org/10.1103/PhysRevD.106.064014}{Phys. Rev. D \textbf{106} (2022) no.6, 064014} ;
\href{https://arxiv.org/pdf/2205.13548}{\tt [arXiv:2205.13548 [gr-qc]]}.
%0 citations counted in INSPIRE as of 26 Jun 2022

\bibitem{marcoa-thermo}
M.~Astorino,
{\it ``Thermodynamics of Regular Accelerating Black Holes''},
\href{https://doi.org/10.1103/PhysRevD.95.064007}{Phys. Rev. D \textbf{95} (2017) no.6, 064007}
\href{https://arxiv.org/pdf/1612.04387.pdf}{\tt [arXiv:1612.04387 [gr-qc]]}
%54 citations counted in INSPIRE as of 22 Jun 2023

\bibitem{marcoa-removal}
M.~Astorino,
{\it ``Removal of conical singularities from rotating C-metrics and dual CFT entropy''}
\href{https://doi.org/10.1007/JHEP10(2022)074}{JHEP \textbf{10} (2022), 074};
\href{https://arxiv.org/pdf/2207.14305}{\tt [arXiv:2207.14305 [gr-qc]]}.

\bibitem{Griffiths:2009dfa}
J.~B.~Griffiths and J.~Podolsky,
{\it ``Exact Space-Times in Einstein's General Relativity''}, 
\href{https://doi.org/10.1017/CBO9780511635397}{Cambridge University Press, 2009}
%doi:10.1017/CBO9780511635397
%135 citations counted in INSPIRE as of 02 May 2023

\bibitem{stephani-big-book}
  H.~Stephani, D.~Kramer, M.~A.~H.~MacCallum, C.~Hoenselaers and E.~Herlt,
  {``Exact solutions of Einstein's field equations''}, \ 
  \href{https://doi.org/10.1017/CBO9780511535185}{\tt [doi:10.1017/CBO9780511535185]}

\bibitem{bubble}
M.~Astorino, R.~Emparan and A.~Vigan\`o,
{\it ``Bubbles of nothing in binary black holes and black rings, and viceversa''},
\href{https://doi.org/10.1007/JHEP07(2022)007}{JHEP \textbf{07} (2022), 007} ;
\href{https://arxiv.org/pdf/2204.09690.pdf}{\tt [arXiv:2204.09690 [hep-th]]}.
%0 citations counted in INSPIRE as of 25 Jul 2022

%\cite{Majumdar:1947eu}
\bibitem{majumdar}
S.~D.~Majumdar,
{\it ``A class of exact solutions of Einstein's field equations''},
\href{https://doi.org/10.1103/PhysRev.72.390}{Phys. Rev. \textbf{72} (1947), 390-398}
%doi:10.1103/PhysRev.72.390
%457 citations counted in INSPIRE as of 20 Jul 2023

\bibitem{belinski-book}
V. Belinski, E. Verdaguer, \href{https://doi.org/10.1017/CBO9780511535253}{\it ``Gravitational solitons''}, Cambridge, Cambridge Univ. Press, 2001.


\bibitem{alekseev-belinski-2RN}
G.~A.~Alekseev and V.~A.~Belinski,
{\it ``Superposition of fields of two Reissner - Nordstrom sources''},
%doi:10.1142/9789812834300_0022
\href{https://arxiv.org/abs/0710.2515}{\tt [arXiv:0710.2515 [gr-qc]]}.
%%CITATION = doi:10.1142/9789812834300_0022;%%
%10 citations counted in INSPIRE as of 03 May 2019


\bibitem{manko-2007}
V.~S.~Manko,
{\it ``The Double-Reissner-Nordstrom solution and the interaction force between two spherically symmetric charged particles''},
Phys.\ Rev.\ D {\bf 76} (2007) 124032;\
%doi:10.1103/PhysRevD.76.124032
\href{https://doi.org/10.1103/PhysRevD.76.124032}{\tt [arXiv:0710.2158 [gr-qc]]}.
%%CITATION = doi:10.1103/PhysRevD.76.124032;%%
%26 citations counted in INSPIRE as of 03 May 2019

\bibitem{compere-kerr-cft}
G.~Comp\`ere,
{\it ``The Kerr/CFT correspondence and its extensions''},
\href{https://doi.org/10.1007/s41114-017-0003-2}{Living Rev. Rel. \textbf{15} (2012), 11} ;
%doi:10.1007/s41114-017-0003-2
\href{https://arxiv.org/pdf/1203.3561}{\tt [arXiv:1203.3561 [hep-th]]}.%[arXiv:1203.3561 [hep-th]].
%197 citations counted in INSPIRE as of 23 Jun 2022

\bibitem{lucietti-kunduri}
H.~K.~Kunduri, J.~Lucietti and H.~S.~Reall,
{\it ``Near-horizon symmetries of extremal black holes''},
\href{https://doi.org/10.1088/0264-9381/24/16/012}{Class. Quant. Grav. \textbf{24} (2007), 4169-4190} ;
%doi:10.1088/0264-9381/24/16/012
\href{https://arxiv.org/pdf/0705.4214}{\tt [arXiv:0705.4214 [hep-th]]}.%[arXiv:0705.4214 [hep-th]].
%282 citations counted in INSPIRE as of 23 Jun 2022

\bibitem{strominger-kerr-cft}
M.~Guica, T.~Hartman, W.~Song and A.~Strominger,
{\it ``The Kerr/CFT Correspondence''},
\href{https://doi.org/10.1103/PhysRevD.80.124008}{Phys. Rev. D \textbf{80} (2009), 124008} ;
%doi:10.1103/PhysRevD.80.124008
\href{https://arxiv.org/pdf/0809.4266}{\tt [arXiv:0809.4266 [hep-th]]}.%[arXiv:0809.4266 [hep-th]].
%722 citations counted in INSPIRE as of 23 Jun 2022

\bibitem{strominger-duals}
T.~Hartman, K.~Murata, T.~Nishioka and A.~Strominger,
{\it ``CFT Duals for Extreme Black Holes''},
\href{https://doi.org/10.1088/1126-6708/2009/04/019}{JHEP \textbf{04} (2009), 019} ;
%doi:10.1088/1126-6708/2009/04/019
\href{https://arxiv.org/pdf/0811.4393}{\tt [arXiv:0811.4393 [hep-th]]}.%[arXiv:0811.4393 [hep-th]].
%219 citations counted in INSPIRE as of 23 Jun 2022

\bibitem{acc-cft}
M.~Astorino,
{\it ``CFT Duals for Accelerating Black Holes''}
\href{https://doi.org/10.1016/j.physletb.2016.07.019}{Phys. Lett. B \textbf{760} (2016), 393-405} ;
%doi:10.1016/j.physletb.2016.07.019
\href{https://arxiv.org/pdf/1605.06131}{\tt [arXiv:1605.06131 [hep-th]]}.
%32 citations counted in INSPIRE as of 23 Jun 2022


\bibitem{melvin-lambda}
M.~Astorino,
{\it ``Charging axisymmetric space-times with cosmological constant''},
\href{https://doi.org/10.1007/JHEP06(2012)086}{JHEP \textbf{06} (2012), 086} ;
\href{https://arxiv.org/pdf/1205.6998.pdf}{\tt [arXiv:1205.6998 [gr-qc]]}.
%35 citations counted in INSPIRE as of 27 Sep 2022


\bibitem{adolfo}
J. Barrientos and A.~Cisterna,
{\it ``Ehlers Transformations as a Tool for Constructing Accelerating NUT Black Holes''},
\href{https://doi.org/10.1103/PhysRevD.108.024059}{Phys. Rev. D \textbf{108} (2023) no.2, 024059}; 
\href{https://arxiv.org/pdf/2305.03765.pdf}{\tt[arXiv:2305.03765 [gr-qc]]}.

\bibitem{marcoa-embedding}
M.~Astorino,
{\it ``Embedding hairy black holes in a magnetic universe''},
\href{https://doi.org/10.1103/PhysRevD.87.084029}{Phys. Rev. D \textbf{87} (2013) no.8, 084029} ;
\href{https://arxiv.org/pdf/1301.6794}{\tt [arXiv:1301.6794 [gr-qc]]}.
%21 citations counted in INSPIRE as of 21 May 2022

\bibitem{marcoa-stationary}
M.~Astorino,
{\it ``Stationary axisymmetric spacetimes with a conformally coupled scalar field''},
\href{http://doi.org/10.1103/PhysRevD.91.064066}{Phys. Rev. D \textbf{91} (2015), 064066} ;
\href{https://arxiv.org/pdf/1412.3539}{\tt [arXiv:1412.3539 [gr-qc]]}.

%\cite{Astorino:2021dju}
\bibitem{marcoa-binary}
M.~Astorino and A.~Vigan\`o,
{\it ``Binary black hole system at equilibrium''},
\href{https://doi.org/10.1016/j.physletb.2021.136506}{Phys. Lett. B \textbf{820} (2021), 136506};
%doi:10.1016/j.physletb.2021.136506
\href{https://arxiv.org/pdf/2104.07686}{\tt [2104.07686 [gr-qc]]}.
%4 citations counted in INSPIRE as of 13 Mar 2022




\bibitem{enhanced}
  M.~Astorino,
  {\it ``Enhanced Ehlers Transformation and the Majumdar-Papapetrou-NUT Spacetime''},
  \href{https://doi.org/10.1007/JHEP01(2020)123}{JHEP \textbf{01} (2020), 123} ; 
   %doi:10.1007/JHEP01(2020)123
   \href{https://arxiv.org/pdf/1906.08228}{\tt [arXiv:1906.08228 [gr-qc]]}.
    %3 citations counted in INSPIRE as of 17 Jan 2022

\bibitem{ernst-remove}
  F.~J.~Ernst,
  {\it ``Removal of the nodal singularity of the C-metric''}, 
   \href{http://scitation.aip.org/content/aip/journal/jmp/17/4/10.1063/1.522935}{J. Math. Phys. {\bf 17}, 515 (1976)}.

\bibitem{ernst-generalized-c} 
  F.~J.~Ernst,
  {\it ``Generalized C-metric''},
   \href{https://doi.org/10.1063/1.523896}{J.\ Math.\ Phys.\  {\bf 19}, 1986-1987 (1978).}

\bibitem{marcoa-pair}
  M.~Astorino,
  {\it ``Pair Creation of Rotating Black Holes''},
  \href{https://doi.org/10.1103/PhysRevD.89.044022}{Phys. Rev. D \textbf{89} (2014) no.4, 044022};
  %doi:10.1103/PhysRevD.89.044022
  \href{https://arxiv.org/pdf/1312.1723.pdf}{\tt [arXiv:1312.1723~[gr-qc]]}.
  %9 citations counted in INSPIRE as of 20 May 2022

\bibitem{multipolar-acc}
  M.~Astorino and A.~Vigan\`o,
  {\it``Many accelerating distorted black holes''},
  \href{https://doi.org/10.1140/epjc/s10052-021-09693-6}{Eur. Phys. J. C \textbf{81} (2021) no.10, 891};
  %doi:10.1140/epjc/s10052-021-09693-6
  \href{https://arxiv.org/pdf/2106.02058.pdf}{\tt [2106.02058 [gr-qc]]}.
  %6 citations counted in INSPIRE as of 05 Apr 2022

\bibitem{misner-counterexample}
  C. W. Misner,  
  {\it ``Taub-NUT space as a counterexample to almost anything''}, 
  Relativity theory and astrophysics 1 (1967): 160.\\  \href{https://ntrs.nasa.gov/api/citations/19660007407/downloads/19660007407.pdf}{\tt [https://ntrs.nasa.gov/api/citations/19660007407/downloads/19660007407.pdf]}

\bibitem{Plebanski-Demianski}
  J.~F.~Plebanski and M.~Demianski,
  {\it ``Rotating, charged, and uniformly accelerating mass in general relativity''},
  \href{https://doi.org/10.1016/0003-4916(76)90240-2}{Annals Phys. \textbf{98} (1976), 98-127}.
  %509 citations counted in INSPIRE as of 18 Jan 2023

\bibitem{Podolsky-2020}
J.~Podolsky and A.~Vratny,
{\it ``Accelerating NUT black holes''}, 
 \href{https://doi.org/10.1103/PhysRevD.102.084024}{Phys. Rev. D \textbf{102} (2020) no.8, 084024}; 
\href{https://arxiv.org/pdf/2007.09169.pdf}{\tt [arXiv:2007.09169 [gr-qc]]}. 
%9 citations counted in INSPIRE as of 18 Jan 2023

\bibitem{mann-stelea-chng}
B.~Chng, R.~B.~Mann and C.~Stelea,
{\it ``Accelerating Taub-NUT and Eguchi-Hanson solitons in four dimensions''},
 \href{https://doi.org/10.1103/PhysRevD.74.084031}{Phys. Rev. D \textbf{74} (2006), 084031} ;
\href{https://arxiv.org/pdf/gr-qc/0608092.pdf}{\tt [arXiv:gr-qc/0608092]}
%12 citations counted in INSPIRE as of 18 Jan 2023

\bibitem{reina-treves}
  A.~Reina and A. Treves
  {\it ``NUT-like generalization of axisymmetric gravitational fields''} ,  \href{https://doi.org/10.1063/1.522614}{Journal of Mathematical Physics 16, 834 (1975)}.

\bibitem{swirling}
M.~Astorino, R.~Martelli and A.~Vigan\`o,
{\it ``Black holes in a swirling universe''},
\href{https://doi.org/10.1103/PhysRevD.106.064014}{Phys. Rev. D \textbf{106} (2022) no.6, 064014} ;
\href{https://arxiv.org/pdf/2205.13548}{\tt [arXiv:2205.13548 [gr-qc]]}.
%0 citations counted in INSPIRE as of 26 Jun 2022

\bibitem{marcoa-removal}
M.~Astorino,
{\it ``Removal of conical singularities from rotating C-metrics and dual CFT entropy''}
\href{https://doi.org/10.1007/JHEP10(2022)074}{JHEP \textbf{10} (2022), 074};
\href{https://arxiv.org/pdf/2207.14305}{\tt [arXiv:2207.14305 [gr-qc]]}.


\bibitem{Podolsky-2021}
J.~Podolsky and A.~Vratny,
{\it ``New improved form of black holes of type D''},
\href{https://doi.org/10.1103/PhysRevD.104.084078}{Phys. Rev. D \textbf{104} (2021), 084078}
%doi:10.1103/PhysRevD.104.084078
\href{https://arxiv.org/pdf/2108.02239}{\tt [arXiv:2108.02239 [gr-qc]]}.
%9 citations counted in INSPIRE as of 27 Mar 2023

\bibitem{Podolsky-2022}
J.~Podolsky and A.~Vratny,
{\it ``New form of all black holes of type D with a cosmological constant''}, 
\href{https://doi.org/10.1103/PhysRevD.107.084034}{Phys. Rev. D \textbf{107} (2023) no.8, 084034}
%doi:10.1103/PhysRevD.107.084034
\href{https://arxiv.org/pdf/2212.08865}{\tt [arXiv:2212.08865 [gr-qc]]}.
%1 citations counted in INSPIRE as of 02 May 2023


\bibitem{bonnor}
W.~B.~Bonnor,
{\it ``A new interpretation of the NUT metric in general relativity''},
\href{https://doi.org/10.1017/s0305004100044807}{Math. Proc. Cambridge Phil. Soc. \textbf{66} (1969) no.1, 145-151}
%42 citations counted in INSPIRE as of 02 May 2023

\bibitem{belinski-book}
V. Belinski, E. Verdaguer, \href{https://doi.org/10.1017/CBO9780511535253}{\it ``Gravitational solitons''}, Cambridge, Cambridge Univ. Press, 2001.

\bibitem{marcoa-embedding}
M.~Astorino,
{\it ``Embedding hairy black holes in a magnetic universe''},
\href{https://doi.org/10.1103/PhysRevD.87.084029}{Phys. Rev. D \textbf{87} (2013) no.8, 084029} ;
\href{https://arxiv.org/pdf/1301.6794}{\tt [arXiv:1301.6794 [gr-qc]]}.
%21 citations counted in INSPIRE as of 21 May 2022

\bibitem{marcoa-stationary}
M.~Astorino,
{\it ``Stationary axisymmetric spacetimes with a conformally coupled scalar field''},
\href{http://doi.org/10.1103/PhysRevD.91.064066}{Phys. Rev. D \textbf{91} (2015), 064066} ;
\href{https://arxiv.org/pdf/1412.3539}{\tt [arXiv:1412.3539 [gr-qc]]}.
%20 citations counted in INSPIRE as of 21 May 2022

\bibitem{alekseev-belinski-kerr}
  G.~A.~Alekseev and V.~A.~Belinski,
  {\it ``Superposition of fields of two rotating charged masses in general relativity and existence of equilibrium configurations''},
  Gen.\ Rel.\ Grav.\  {\bf 51} (2019) no.5,  68
  %doi:10.1007/s10714-019-2543-0
  \href{https://arxiv.org/abs/1905.05317}{\tt [arXiv:1905.05317 [gr-qc]]}.
  %%CITATION = doi:10.1007/s10714-019-2543-0;%%
  %1 citations counted in INSPIRE as of 22 Oct 2019

\bibitem{bubble}
M.~Astorino, R.~Emparan and A.~Vigan\`o,
{\it ``Bubbles of nothing in binary black holes and black rings, and viceversa''},
\href{https://doi.org/10.1007/JHEP07(2022)007}{JHEP \textbf{07} (2022), 007} ;
\href{https://arxiv.org/pdf/2204.09690.pdf}{\tt [arXiv:2204.09690 [hep-th]]}.
%0 citations counted in INSPIRE as of 25 Jul 2022

\bibitem{tesi-giova}
G.~Boldi,
{\it ``Ehlers transformation and accelerating spacetimes with a gravomagnetic monopole''},
\href{https://inspirehep.net/literature/2652409}{\tt Università degli Studi di Milano (2022)}

\bibitem{asto-lambda}
M.~Astorino,
{\it ``Charging axisymmetric space-times with cosmological constant''},
\href{https://doi.org/10.1007/JHEP06(2012)086}{JHEP \textbf{06} (2012), 086} ;
\href{https://arxiv.org/pdf/1205.6998.pdf}{\tt [arXiv:1205.6998 [gr-qc]]}.
%35 citations counted in INSPIRE as of 27 Sep 2022

\bibitem{adolfo}
J. Barrientos and A.~Cisterna,
{\it ``Ehlers Transformations as a Tool for Constructing Accelerating NUT Black Holes''},
\href{https://arxiv.org/pdf/2305.03765.pdf}{\tt[arXiv:2305.03765 [gr-qc]]}.
%0 citations counted in INSPIRE as of 16 May 2023{\it ``Ehlers Transformations as a Tool for Constructing Accelerating NUT Black Holes''}, to appear soon.


\bibitem{ernst-remove}
  F. Ernst
  {\it ``Removal of the nodal singularity of the C-metric''}, 
   \href{http://scitation.aip.org/content/aip/journal/jmp/17/4/10.1063/1.522935}{J. Math. Phys. {\bf 17}, 515 (1976)}.


\bibitem{marcoa-pair}
M.~Astorino,
{\it ``Pair Creation of Rotating Black Holes''},
\href{https://doi.org/10.1103/PhysRevD.89.044022}{Phys. Rev. D \textbf{89} (2014) no.4, 044022};
%doi:10.1103/PhysRevD.89.044022
\href{https://arxiv.org/pdf/1312.1723.pdf}{\tt [arXiv:1312.1723~[gr-qc]]}.
%9 citations counted in INSPIRE as of 20 May 2022

\bibitem{ernst-magnetic} 
  F.~J.~Ernst,
  {\it ``Black holes in a magnetic universe''},
  \href{https://doi.org/10.1063/1.522781}{J.\ Math.\ Phys.\  {\bf 17}, no. 1, 54 (1976).}


\bibitem{ernst-generalized-c} 
  F.~J.~Ernst,
  {\it ``Generalized C-metric''},
   \href{https://doi.org/10.1063/1.523896}{J.\ Math.\ Phys.\  {\bf 19}, 1986-1987 (1978).}


%\cite{Astorino:2021dju}
\bibitem{marcoa-binary}
M.~Astorino and A.~Vigan\`o,
{\it ``Binary black hole system at equilibrium''},
\href{https://doi.org/10.1016/j.physletb.2021.136506}{Phys. Lett. B \textbf{820} (2021), 136506};
%doi:10.1016/j.physletb.2021.136506
\href{https://arxiv.org/pdf/2104.07686}{\tt [2104.07686 [gr-qc]]}.
%4 citations counted in INSPIRE as of 13 Mar 2022


\bibitem{deCastro}
G.~M.~de Castro and P.~S.~Letelier,
{\it ``Black holes surrounded by thin rings and the stability of circular orbits''},
\href{https://doi.org/10.1088/0264-9381/28/22/225020}{Class. Quant. Grav. \textbf{28} (2011), 225020}

%\cite{Astorino:2021rdg}
\bibitem{multipolar-acc}
M.~Astorino and A.~Vigan\`o,
{\it``Many accelerating distorted black holes''},
\href{https://doi.org/10.1140/epjc/s10052-021-09693-6}{Eur. Phys. J. C \textbf{81} (2021) no.10, 891};
%doi:10.1140/epjc/s10052-021-09693-6
\href{https://arxiv.org/pdf/2106.02058.pdf}{\tt [2106.02058 [gr-qc]]}.
%6 citations counted in INSPIRE as of 05 Apr 2022

\bibitem{swirling}
M.~Astorino, R.~Martelli and A.~Vigan\`o,
{\it ``Black holes in a swirling universe''},
\href{https://arxiv.org/pdf/2205.13548}{\tt [arXiv:2205.13548 [gr-qc]]}.
%0 citations counted in INSPIRE as of 26 Jun 2022

\bibitem{reina-treves}
  A.~Reina and A. Treves
  {\it ``NUT-like generalization of axisymmetric gravitational fields''},
  \href{https://doi.org/10.1063/1.522614}{Journal of Mathematical Physics 16, 834 (1975)}.

\bibitem{enhanced}
M.~Astorino,
{\it ``Enhanced Ehlers Transformation and the Majumdar-Papapetrou-NUT Spacetime''},
\href{https://doi.org/10.1007/JHEP01(2020)123}{JHEP \textbf{01} (2020), 123} ; 
%doi:10.1007/JHEP01(2020)123
\href{https://arxiv.org/pdf/1906.08228}{\tt [arXiv:1906.08228 [gr-qc]]}.
%3 citations counted in INSPIRE as of 17 Jan 2022


\bibitem{marcoa-embedding}
M.~Astorino,
{``Embedding hairy black holes in a magnetic universe''},
\href{https://doi.org/10.1103/PhysRevD.92.104006}{Phys. Rev. D \textbf{92} (2015) no.10, 104006};
\href{https://doi.org/10.1103/PhysRevD.87.084029}{Phys. Rev. D \textbf{87} (2013) no.8, 084029} ;
\href{https://arxiv.org/pdf/1301.6794}{\tt [arXiv:1301.6794 [gr-qc]]}.
%21 citations counted in INSPIRE as of 21 May 2022

\bibitem{marcoa-stationary}
M.~Astorino,
{\it ``Stationary axisymmetric spacetimes with a conformally coupled scalar field''},
\href{http://doi.org/10.1103/PhysRevD.91.064066}{Phys. Rev. D \textbf{91} (2015), 064066} ;
\href{https://arxiv.org/pdf/1412.3539}{\tt [arXiv:1412.3539 [gr-qc]]}.
%20 citations counted in INSPIRE as of 21 May 2022

\bibitem{magnetised-kerr-cft}
M.~Astorino,
{ \it ``Magnetised Kerr/CFT correspondence''},
\href{https://doi.org/10.1016/j.physletb.2015.10.017}{Phys. Lett. B \textbf{751} (2015), 96-106};
\href{https://arxiv.org/pdf/1508.01583}{\tt [arXiv:1508.01583 [hep-th]]}.
%31 citations counted in INSPIRE as of 23 Jun 2022

\bibitem{magnetised-RN-cft}
M.~Astorino,
{\it ``Microscopic Entropy of the Magnetised Extremal Reissner-Nordstrom Black Hole''},
\href{https://doi.org/10.1007/JHEP10(2015)016}{JHEP \textbf{10} (2015), 016} ;
\href{https://arxiv.org/pdf/1507.04347}{\tt [arXiv:1507.04347 [hep-th]]}.
%18 citations counted in INSPIRE as of 23 Jun 2022


\bibitem{c-cft}
M.~Astorino,
{\it ``CFT Duals for Accelerating Black Holes''}
\href{https://doi.org/10.1016/j.physletb.2016.07.019}{Phys. Lett. B \textbf{760} (2016), 393-405} ;
%doi:10.1016/j.physletb.2016.07.019
\href{https://arxiv.org/pdf/1605.06131}{\tt [arXiv:1605.06131 [hep-th]]}.
%32 citations counted in INSPIRE as of 23 Jun 2022


\bibitem{lucietti-kunduri}
H.~K.~Kunduri, J.~Lucietti and H.~S.~Reall,
{\it ``Near-horizon symmetries of extremal black holes''},
\href{https://doi.org/10.1088/0264-9381/24/16/012}{Class. Quant. Grav. \textbf{24} (2007), 4169-4190} ;
%doi:10.1088/0264-9381/24/16/012
\href{https://arxiv.org/pdf/0705.4214}{\tt [arXiv:0705.4214 [hep-th]]}.%[arXiv:0705.4214 [hep-th]].
%282 citations counted in INSPIRE as of 23 Jun 2022

\bibitem{strominger-kerr-cft}
M.~Guica, T.~Hartman, W.~Song and A.~Strominger,
{\it ``The Kerr/CFT Correspondence''},
\href{https://doi.org/10.1103/PhysRevD.80.124008}{Phys. Rev. D \textbf{80} (2009), 124008} ;
%doi:10.1103/PhysRevD.80.124008
\href{https://arxiv.org/pdf/0809.4266}{\tt [arXiv:0809.4266 [hep-th]]}.%[arXiv:0809.4266 [hep-th]].
%722 citations counted in INSPIRE as of 23 Jun 2022

\bibitem{strominger-duals}
T.~Hartman, K.~Murata, T.~Nishioka and A.~Strominger,
{\it ``CFT Duals for Extreme Black Holes''},
\href{https://doi.org/10.1088/1126-6708/2009/04/019}{JHEP \textbf{04} (2009), 019} ;
%doi:10.1088/1126-6708/2009/04/019
\href{https://arxiv.org/pdf/0811.4393}{\tt [arXiv:0811.4393 [hep-th]]}.%[arXiv:0811.4393 [hep-th]].
%219 citations counted in INSPIRE as of 23 Jun 2022

\bibitem{compere-kerr-cft}
G.~Comp\`ere,
{\it ``The Kerr/CFT correspondence and its extensions''},
\href{https://doi.org/10.1007/s41114-017-0003-2}{Living Rev. Rel. \textbf{15} (2012), 11} ;
%doi:10.1007/s41114-017-0003-2
\href{https://arxiv.org/pdf/1203.3561}{\tt [arXiv:1203.3561 [hep-th]]}.%[arXiv:1203.3561 [hep-th]].
%197 citations counted in INSPIRE as of 23 Jun 2022

\bibitem{gaston}
L.~Donnay, G.~Giribet, H.~A.~Gonz\'alez and M.~Pino,
{\it ``Extended Symmetries at the Black Hole Horizon''},
\href{https://doi.org/10.1007/JHEP09(2016)100}{JHEP \textbf{09} (2016), 100} ;
%doi:10.1007/JHEP09(2016)100
\href{https://arxiv.org/pdf/1607.05703}{\tt [arXiv:1607.05703 [hep-th]]}.%[arXiv:1607.05703 [hep-th]].
%132 citations counted in INSPIRE as of 24 Jun 2022

\bibitem{bicak}
J.~Bi\v{c}\'ak and F.~Hejda,
{\it ``Near-horizon description of extremal magnetized stationary black holes and Meissner effect''},
\href{https://doi.org/10.1103/PhysRevD.92.104006}{Phys. Rev. D \textbf{92} (2015) no.10, 104006}; 
\href{https://arxiv.org/pdf/1510.01911.pdf}{\tt [arXiv:1510.01911 [gr-qc]]}.
%28 citations counted in INSPIRE as of 24 Jun 2022

\bibitem{ernst2}
F.~J.~Ernst,
  {\it ``New Formulation of the Axially Symmetric Gravitational Field Problem. II''},
  \href{https://doi.org/10.1103/PhysRev.168.1415}{Phys.\ Rev.\  {\bf 168} (1968) 1415}.

\bibitem{strominger}
D.~Garfinkle, S.~B.~Giddings and A.~Strominger,
{\it ``Entropy in black hole pair production''},
\href{https://doi.org/10.1103/PhysRevD.49.958}{Phys. Rev. D \textbf{49} (1994), 958-965};
%doi:10.1103/PhysRevD.49.958
\href{https://arxiv.org/pdf/gr-qc/9306023.pdf}{\tt [arXiv:gr-qc/9306023]}
%116 citations counted in INSPIRE as of 20 May 2022

\bibitem{hawking}
S.~W.~Hawking, G.~T.~Horowitz and S.~F.~Ross,
{\it ``Entropy, Area, and black hole pairs''},
\href{https://doi.org/10.1103/PhysRevD.51.4302}{Phys. Rev. D \textbf{51} (1995), 4302-4314};
%doi:10.1103/PhysRevD.51.4302
\href{https://arxiv.org/pdf/gr-qc/9409013.pdf}{\tt [arXiv:gr-qc/9409013]}.
%389 citations counted in INSPIRE as of 20 May 2022

\bibitem{KramerNeug}
D.~Kramer, G.~Neugebauer,
{\it ``The superposition of two Kerr solutions''},
\href{https://www.sciencedirect.com/science/article/abs/pii/0375960180905563}{Physics Letters A \textbf{75} (1980), no.4, 259--261}.

\bibitem{bubble}
M.~Astorino, R.~Emparan and A.~Vigan\`o,
{\it ``Bubbles of nothing in binary black holes and black rings, and viceversa''},
\href{https://doi.org/10.1007/JHEP07(2022)007}{JHEP \textbf{07} (2022), 007} ;
\href{https://arxiv.org/pdf/2204.09690.pdf}{\tt [arXiv:2204.09690 [hep-th]]}.
%0 citations counted in INSPIRE as of 25 Jul 2022


\bibitem{asto-lambda}
M.~Astorino,
{\it ``Charging axisymmetric space-times with cosmological constant''},
\href{https://doi.org/10.1007/JHEP06(2012)086}{JHEP \textbf{06} (2012), 086} ;
\href{https://arxiv.org/pdf/1205.6998.pdf}{\tt [arXiv:1205.6998 [gr-qc]]}.
%35 citations counted in INSPIRE as of 27 Sep 2022


\bibitem{kerr-magnetic}  
  F.~J.~Ernst and W. Wild,
  {\it ``Kerr black holes in a magnetic universe''},
   J.\ Math.\ Phys.{\bf 17} (1976) 182.


\bibitem{misner-couterexample}
C. W. Misner,  
{\it Taub-NUT space as a counterexample to almost anything},
Relativity theory and astrophysics 1 (1967): 160.\\
\href{https://ntrs.nasa.gov/api/citations/19660007407/downloads/19660007407.pdf}{\tt [https://ntrs.nasa.gov/api/citations/19660007407/downloads/19660007407.pdf]}


 \bibitem{geroch1}
  R.~P.~Geroch,
  {\it ``A Method for generating solutions of Einstein's equations''},
  J.\ Math.\ Phys.\  {\bf 12} (1971) 918.
  %%CITATION = JMAPA,12,918;%%
  %310 citations counted in INSPIRE as of 27 May 2014
  
  \bibitem{geroch2}
  R.~P.~Geroch,
  {\it ``A Method for generating new solutions of Einstein's equation. 2''},
  J.\ Math.\ Phys.\  {\bf 13} (1972) 394.
  %%CITATION = JMAPA,13,394;%%
  %227 citations counted in INSPIRE as of 27 May 2014

\bibitem{hauser-ernst}
I.~Hauser and F.~J.~Ernst,
{\it ``Proof of a generalized Geroch conjecture for the hyperbolic Ernst equation''},
Gen. Rel. Grav. \textbf{33} (2001), 195-293,
%\href{https://doi.org/10.1023/A:1002701301339}{doi:10.1023/A:1002701301339}
\href{https://arxiv.org/abs/gr-qc/0002049}{\tt[arXiv:gr-qc/0002049]}.
 
 
\bibitem{ernst1}
   F.~J.~Ernst,
  {\it ``New formulation of the axially symmetric gravitational field problem''},
  Phys.\ Rev.\  {\bf 167} (1968) 1175.

\bibitem{ernst2}
F.~J.~Ernst,
  {\it ``New Formulation of the Axially Symmetric Gravitational Field Problem. II''},
  Phys.\ Rev.\  {\bf 168} (1968) 1415. \href{https://doi.org/10.1103/PhysRev.168.1415}{\tt [doi:10.1103/PhysRev.168.1415]}

\bibitem{belinski}
V. Belinski, E. Verdaguer, {\it ``Gravitational solitons''}, Cambridge, Cambridge Univ. Press, 2001.



\bibitem{stephani}
  H.~Stephani, D.~Kramer, M.~A.~H.~MacCallum, C.~Hoenselaers and E.~Herlt,
  {``Exact solutions of Einstein's field equations''}, \ 
  \href{https://doi.org/10.1017/CBO9780511535185}{\tt [doi:10.1017/CBO9780511535185]}


\fi

\end{thebibliography}
\end{document}